\documentclass[]{article}
\usepackage{url}
\usepackage{amssymb, amsfonts, amsmath, amsthm}
\usepackage{graphicx}
\usepackage{xcolor}
\usepackage[colorlinks=true,
	linkcolor=violet,
	citecolor=black]{hyperref}
\usepackage{authblk}
\usepackage{enumitem}
\usepackage{subfig}
\usepackage{apacite}
\usepackage[margin=1in]{geometry}
\newcommand{\BIT}{\begin{itemize}}
\newcommand{\EIT}{\end{itemize}}
\newcommand\mbb[1]{\mathbb{#1}}

\def\reals{\mathbb{R}} % Real number symbol
\renewcommand{\exp}[1]{\operatorname{exp}\left(#1\right)} % Exponential
\def\indic#1{\mbb{I}\left({#1}\right)} % Indicator function
\def\absarg#1{\left|#1\right|}

\def\Gsn{\mathcal{N}}

\def\Haar{\textnormal{Haar}}
\theoremstyle{definition}

\def\*#1{\mathbf{#1}}

\graphicspath{{figure/}}

\title{Bootstrap Confidence Regions for Learned Feature Embeddings}
\author{Kris Sankaran}
\affil{Department of Statistics \\ University of Wisconsin - Madison \\ \href{mailto:ksankaran@wisc.edu}{ksankaran@wisc.edu}}

\begin{document}
\maketitle

\begin{abstract}
    Algorithmic feature learners provide high-dimensional vector representations for non-matrix structured signals, like images, audio, text, and graphs. Low-dimensional projections derived from these representations can be used to explore variation across collections of these data. However, it is not clear how to assess the uncertainty associated with these projections. We adapt methods developed for bootstrapping principal components analysis to the setting where features are learned from non-matrix data. We empirically compare the derived confidence regions in simulations, varying factors that influence both feature learning and the bootstrap. Approaches are illustrated on spatial proteomic data. Code, data, and trained models are released as an R compendium.
\end{abstract}

Dimensionality reduction is often used to explore the high-dimensional representations made by feature learning algorithms \cite{erhan2009visualizing, nguyen2019understanding}. For example, principal components analysis (PCA) applied to word embedding vectors can highlight documents with similar content. These projections are typically treated as fixed, and their uncertainty is rarely characterized. A separate line of work, predating the widespread use of algorithmic feature learners, investigated bootstrap strategies for visualizing the uncertainty associated with dimensionality reduction methods \cite{josse2016confidence, chateau1996assessing, elguero1988confidence}. The purpose of this study is to adapt these bootstrap methods to the algorithmic feature learning setting, accounting for difficulties that
arise when using the output of one model as input to another \cite{wang2020methods, chernozhukov2017double}.

For example, in the word embedding application above, training the embedding model twice may result in two different (though hopefully similar) representations of a document corpus. This introduces randomness in the feature extraction process that was not present when studying data-induced uncertainty  in deterministic methods. Furthermore, the coordinates of the two learned embeddings need not correspond to one another. This is different from the usual setting where, though new rows may be sampled from the population, the columns of the data matrix are assumed fixed. Moreover, many feature learning algorithms are supervised, and using labels to develop representations that are predictive of a response. If the model is overfit, the associated dimensionality reduction may give misleading interpretations, overstating the differences between classes in the learned feature space \cite{gross2015selective, friedman2001elements}.

For these reasons, visualizing uncertainty in learned feature embeddings requires additional care. We make the following contributions,

\begin{itemize}
    \item Computing confidence areas: We describe three bootstrap-based strategies for computing 
    confidence areas on dimensionality-reduced learned features. The approaches differ
    in the assumptions they make and the resources they require, and we clarify the effects of these choices
    on downstream analysis.
    \item Simulation and data analysis test-bed: We design generative mechanisms with enough complexity to warrant the application of feature learning algorithms, but which are transparent enough to support analysis. We prepare a dataset for a spatial proteomics problem where characterizing projection uncertainty has practical implications.
    \item Feature learner and dataset effects: We train supervised, unsupervised, or random-feature based learners using varying amounts of data and across degrees of model complexity. This clarifies whether projections from a supervised Convolutional Neural Network (CNN) require different treatment than those from an unsupervised Variational Autoencoder (VAE), for example.
\end{itemize}

Though all bootstrap approaches give comparable results and are approximately valid in a simple low-rank model, more complex simulations suggest that 
there is no universally applicable approach. In more complex contexts, an approach that only trains one feature learner appears to underestimate projection uncertainty across all
types of feature extractors.

Section \ref{sec:background} summarizes technical tools used below. We
present methods for constructing bootstrap confidence regions in Section
\ref{sec:methods}, and we study its properties through simulation in
Section \ref{sec:simulation}. We compare methods in Section 
\ref{sec:data_analysis} through an application to a spatial proteomics dataset. We
summarize in Section \ref{sec:discussion}. We have released all code, raw and 
intermediate data, and trained models associated with simulations and data analysis. 
Documentation about these artifacts and how to reproduce them is provided in 
Supplementary Section \ref{sec:reproducibility}. 

\section{Background}
\label{sec:background}

\subsection{Feature Learning}
\label{subsec:feature_learning}

We consider three complementary feature learning algorithms. The
first is the VAE, which learns an $L$-dimensional reduction of a dataset by
optimizing a reconstruction objective. It posits a
generative model $p\left(z\right)p_{\xi}\left(x \vert z\right)$ of the data;
$p\left(z\right)$ is a prior on latent features and $p_{\xi}\left(x \vert
z\right)$ is a likelihood parameterized by $\xi$. The algorithm finds a pair
$\xi, \varphi$ maximizing the lower bound,
\begin{align*}
\log p_{\xi}\left(x\right) \geq  \mathbb{E}_{q_{\varphi}}\left[\log p_{\xi}(x \mid z)\right]-D_{KL}\left(q_{\varphi}(z \mid x) \| p(z)\right)
\end{align*}
where $q_{\varphi}\left(z \vert x\right) = \Gsn\left(\mu_{\varphi}\left(x\right),
\sigma^{2}_{\varphi}\left(x\right)\right)$ maps raw data examples to
distributions in a latent space. This  problem is nonconvex, and the 
solution is non-deterministic. There are many implementations of VAEs; our
experiments follow \cite{van2017neural}.

Second, we learn supervised features through a
CNN. A CNN regressor optimizes an
empirical estimate of $\mathbf{E}\|y -
f_{W_{1:J}}\left(x\right)^{T}\beta\|_{2}^{2}$ over $W_{1:J}$ and $\beta$. Here,
$f_{W_{1:J}}$ transforms the raw input into the ``final layer'' features, and
is defined recursively according to
\begin{align*}
f^{j}_{W_{1:j}}\left(x\right) &= \sigma\left(W_{j}f^{j - 1}_{W_{1:(j - 1)}}\left(x\right)\right)\\
f^{0}\left(x\right) &= x
\end{align*}
where $\sigma\left(x\right) := x \indic{x \geq
  0}$ and matrices $W$ are restricted to the set of convolutions. Like in
the VAE, this solved through first-order optimization methods. Our implementation is 
the CBR architecture from \cite{raghu2017svcca}.

Third, we use a random convolutional features (RCF) model
\cite{rahimi2008weighted}. A random sample of $L$ training examples
$x_{i_1}, \dots, x_{i_L} \in \reals^{w \times h \times c}$ is selected; the
$x_{i}$'s are assumed to be $c$-channel images with dimension $w\times h$. For
each sample, a random $s \times s$ patch, denoted $w_{p} \in \reals^{s
  \times s \times c}$, is extracted. For any $c$-channel image $x$, the $l^{th}$
feature $z_{l}$ is found by convolving $x$ with $w_{l}$ and spatially averaging
over activations. This model uses random training image patches as
convolutional kernels, rather than learning them from scratch. The features
$z_{1}, \dots, z_{L}$ are analogous to the features $f_{W_{1:J}}\left(x\right)$
in the CNN.

To train an RCF, the training data are featurized into $\*Z \in
\reals^{n \times L}$. Then, a ridge regression model is trained from $\*Z$ 
to the $y$, giving an  estimate $\hat{\beta}$. For a 
new example $x^{\ast}$, the same image patches $w_{1}, \dots,
w_{L}$ are used to form $z^{\ast}$, and
predictions are made with $z^{\ast T}\hat{\beta}$. This model does not require gradient based training, and it can serve as a fast baseline.

\subsection{Procrustes Analysis}
\label{subsec:procrustes}

Given centered $\mathbf{X}$ and $\mathbf{Y}$, the Procrustes problem finds a rotation
$\mathbf{R}$ solving,
\begin{align*}
\min_{\mathbf{R} \in \mathcal{O}\left(p, p\right)} \|\mathbf{X} - \mathbf{Y}\mathbf{R}\|^{2}_{F},
\end{align*}
where $\mathcal{O}\left(p, p\right)$ is the space of $p\times p$
orthonormal matrices. The solution can be shown to be $\hat{\mathbf{R}} = \mathbf{U}^{T}\mathbf{V}$
for $\mathbf{U}$ and $\mathbf{V}$ obtained by the SVD of
$\mathbf{X}^{T}\mathbf{Y}$ \cite{friedman2001elements, gower1975generalized}. For $B$ matrices $\mathbf{X}_{1}, \dots,
\mathbf{X}_{B}$, the generalized Procrustes problem finds $B$ rotations
$\mathbf{R}_{1}, \dots, \mathbf{R}_{B}$ and mean $\mathbf{M}$ solving
\begin{align*}
\min_{\mathbf{R}_{1}, \dots, \mathbf{R}_{B} \in \mathcal{O}\left(p, p\right), M} \sum_{b = 1}^{B} \|\mathbf{X}_{b}\mathbf{R}_{b} - \mathbf{M}\|_{F}^{2}.
\end{align*}
While there is no closed form solution, the optimization can be solved by
cyclically updating each $\mathbf{R}_{b}$ via standard Procrustes problems and
then updating $\mathbf{M} = \frac{1}{B} \sum_{b = 1}^{B} \mathbf{X}_{b}
\mathbf{R}_{b}$ \cite{friedman2001elements}.

\subsection{PCA and the Bootstrap}
\label{subsec:pca_bootstrap}

Several approaches are available for bootstrapping PCA.
The total bootstrap computes $B$ principal planes by applying PCA to $B$ resampled versions of the data \cite{chateau1996assessing}. For each replication, rows are sampled with replacement, viewed as draws from a larger population. The associated principal axes may be reflected or swapped with one another, so the associated sample coordinates are not directly comparable, and coordinates must be aligned. This is often accomplished through a Procrustes or conjoint analysis \cite{elguero1988confidence}. In either case, the cloud of $B$ points associated with each sample in the resulting reference space is used to form a confidence region for it.

In contrast, fixed-effects PCA views the rows of the data matrix as the entire population of interest \cite{josse2016confidence}. The source of randomness in this case is measurement noise around a low-rank model, not sampling from a larger population of rows. By fitting a measurement noise model and resampling residuals, a parametric bootstrap provides confidence regions for the true latent coordinates in the low-rank model. We also note that Bayesian factor analysis approaches sampling from the posterior of latent coordinates \cite{ren2017bayesian, ren2020bayesian}. Like the fixed-effects PCA model, these approaches specify an explicit low-rank model with measurement noise. Like the total bootstrap, underlying factors may be swapped, and alignment is necessary.

\section{Methods}
\label{sec:methods}

This section describes ways to adapt the bootstrap approaches above to the feature learning context.
Our raw data are $n$ samples $x_i \in \mathcal{X}$, where $\mathcal{X}$ is the raw data domain, e.g., images, text sentences, or audio signals. A corresponding set $y_i in \reals$ of responses may be available. The full data are $\mathcal{D} = \left(x_i, y_i\right)_{i = 1}^{n}$.
A \textit{feature learner} is a parameterized mapping $T\left(\cdot;
\theta\right): \mathcal{X} \to \reals^{L}$ taking data from
$\mathcal{X}$ and representing it in $\reals^{L}$.

For example, in a text data application, we expect the learner to transform a set of raw word sequences into a vector of features reflecting the topic of the document. $\theta$ is estimated from data, typically through an optimization,
\begin{align}
\label{eq:optim}
  \hat{\theta} := \arg\min_{\theta \in \Theta} \mathcal{L}\left(\mathcal{D}, T\left(\cdot; \theta\right)\right)
\end{align}
for some loss $\mathcal{L}$. In an unsupervised feature learner,
candidates $\theta \in \Theta$ are functions of $x_{1}, \dots, x_{n}$ alone. For
a supervised feature learner, the class includes functions of both
$x_{1}, \dots, x_{n}$ and $y$. To simplify notation, we will write
$z_{i} = T\left(x_{i}; \hat{\theta}\right) \in
\reals^{L}$ to denote the learned features for observation $i$.

A challenge is that the learned features are not the same from one run to the next; the $l^{th}$ learned feature from run 1 need not have any relationship with the $l^{th}$ feature from run 2. This is a consequence of using stochastic optimization in \ref{eq:optim}. However, even if there is no direct correspondence 
across runs, they may all reflect the same underlying latent
features. 
In particular, projections from dimensionality reductions from the two datasets may be similar, after applying an appropriate alignment. 

Suppose that data have been split into a feature learning set, indexed by $I \subset \{1, \dots, n\}$
and an inference set, indexed by $I^{C}$. The fraction $\frac{1}{n}\absarg{I}$ used for feature
learning is a hyperparameter whose influence is empirically studied below. The learning set $\left(x_{i}\right)_{i \in I}$ is resampled $B$ times, leading to $B$ different feature extractors,
$T\left(\cdot; \hat{\theta}^{b}\right)$ which can then be applied to the full dataset, yielding learned features $\*Z_{b} \in \reals^{n \times L}$

\subsection{Nonparametric Bootstrap}
\label{subsec:nonparametric_bootstrap}

Like the total bootstrap, one approach to comparing embeddings across feature extractors is to perform a dimensionality reduction on each
and then align the projections. For each $b$, compute a singular value decomposition,
\begin{align*}
\*Z_{b, I^{C}} &= \hat{\*U}_{b}\hat{\*\Sigma}_{b}\hat{\*V}_{b}^{T}
\end{align*}
where the index $I^{C}$ means that only features associated with the inference set are used. Define coordinates for sample $i$ with respect to the top $K$ right singular vectors using
$l_{i}^{b} = \sum_{k = 1}^{K} \hat{u}^{b}_{ik}\hat{\sigma}_{k}^{b}$. These can be stacked into $\*L_{b} \in \reals^{\absarg{I^{C}} \times K}$.

A Procrustes analysis applied to $\*L_{1}, \dots, \*L_{B}$ learns a series of rotation matrices $\*R_{1}, \dots, \*R_{B}$ aligning the projections.
For each sample $i$, compute a mean and covariance matrix based on the $B$ vectors $\*R_{b}l_{i}^{b}$.
These are used to create $1-\alpha$ level confidence areas for each inference sample in the $K$-dimensional projection space. 
This approach plugs-in an estimate for a ``true'' low-dimensional $\*L$, assuming that this representation is noisily observed and then subject to arbitrary rotations. Note that if the true latent representations are subject to more general transformations (e.g., translations) across runs
of the extractor, this assumption may not be appropriate.

The advantage of this approach is that it does not require a parametric model for simulating new versions of $\*Z_{b}$. The price to pay is that it is necessary to train $B$ feature extraction models $T\left(\cdot, \hat{\theta}_{b}\right)$, which can be a computational burden, even if it is parallelizable.
Further, confidence areas are not computed for samples in the feature learning
set $\left(x_{i}\right)_{i \in I}$. However,
if the uncertainty of sample-level projections is assumed to vary smoothly, then
a heuristic is to consider the uncertainty of a sample in $I$ as comparable to 
those of its nearby samples in $I^{C}$.

\begin{figure}
    \centering
    \includegraphics[width=1\textwidth]{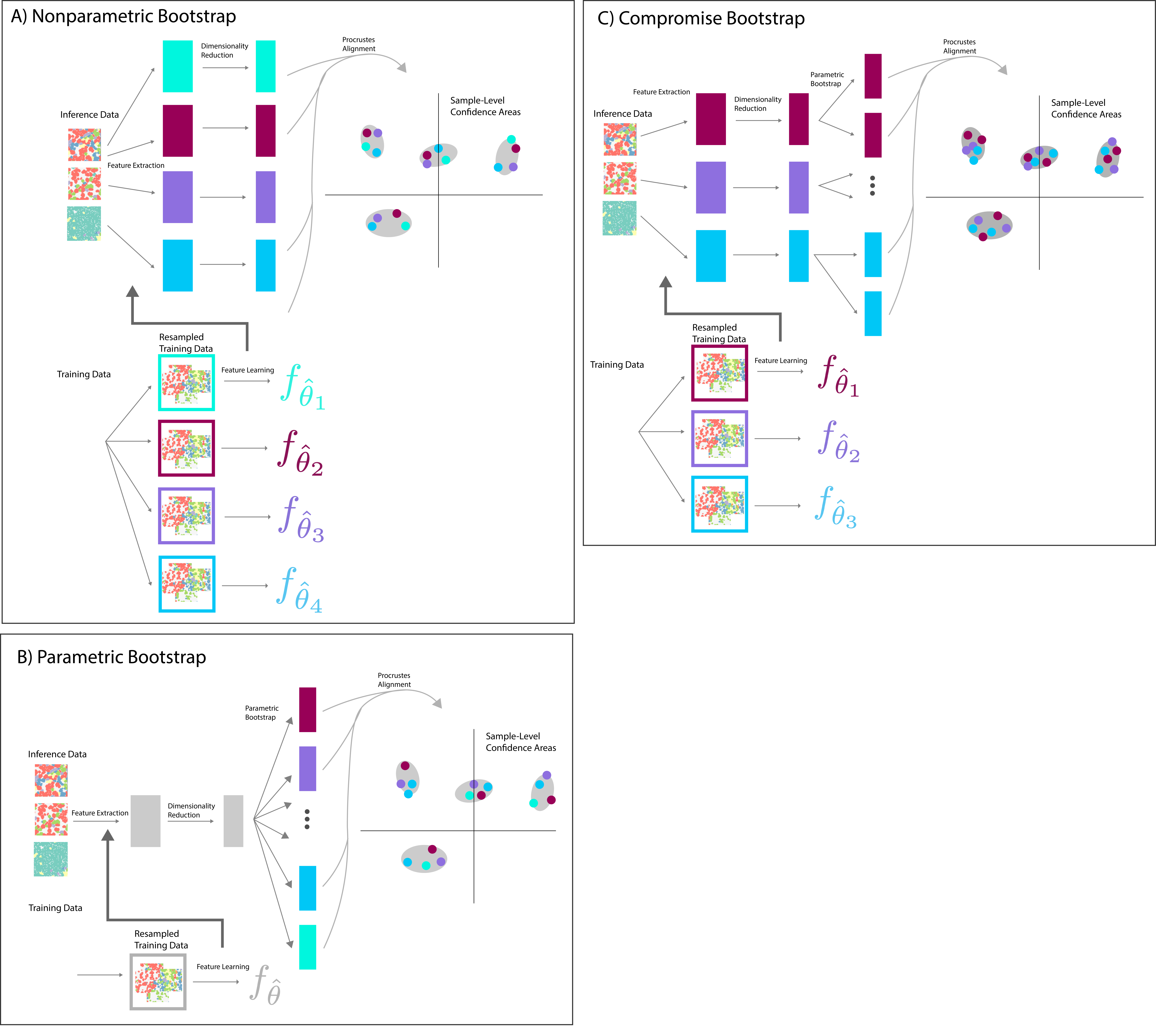}
    \caption{A summary of the proposed bootstrap procedures. All begin by splitting data into training and inference sets, used for feature learning and confidence region construction, respectively. The nonparametric bootstrap (a) trains $B$ separate feature learners, each of which are used for feature extraction and dimensionality reduction before being aligned. The parametric bootstrap (b) trains a single feature learner and then simulates and aligns an ensemble of $B$ latent coordinates for each sample. The compromise (c) trains a smaller set of feature learners but further resamples residuals (like in the parametric bootstrap) to increase the number of apparent bootstrap replicates.}
    \label{fig:combined_summary_graphic}
\end{figure}

\subsection{Parametric Bootstrap}
\label{subsec:parametric_bootstrap}

To avoid the computational complexity associated with training $B$ feature extractors, we consider
a parametric bootstrap, which simulates $\*Z_{b}$ by resampling residuals from an fitted low-rank model, analogous to the fixed-effects PCA approach \cite{josse2016confidence}. Suppose that variation across $x_{i}$ is induced by latent features $l_{i} \in
\reals^{K}$. The feature learning process is modeled as,
\begin{align}
\label{eq:para_boot}
\*Z &= \*L \*V^{T} + \*E & E_{ij} &\sim \Gsn\left(0, \sigma_{\*E}^2\right) \\
y &= \*L \beta + \epsilon & \epsilon_{i} &\sim \Gsn\left(0, \sigma_{\epsilon}^2\right)
\end{align}
where $\*L \in \reals^{n \times K}$ stacks the $l_i$  and $E_{ij}$ is the $ij^{th}$ element of $\*E$. Only $\*Z$ is available for predicting the response $y$.

To simulate $\*Z_{b}$ based on a single set of observed latent features $\*Z$, we resample rows of $\*Z$ in the inference set $I^{C}$ and compute the associated rank-$K$ truncated SVD,
$\hat{\*U}\hat{\*\Sigma}\hat{\*V}^T$.
Then we draw,
\begin{align*}
\*Z_{b} = \left(\hat{\*U}\hat{\*\Sigma} + \*E_{b}\right)\*\Pi_{b},
\end{align*}
where $\*E_{b} \in \reals^{n \times K}$ is obtained by resampling entries of $\*Z - \hat{\*Z}$ and $\*\Pi_{b} \in \reals^{n \times K}$ is a random permutation matrix, reflecting the fact that coordinates of the feature extractor need not match from one run to the next. Alignment and confidence area construction then proceeds as in section \ref{subsec:nonparametric_bootstrap}.

\subsection{Compromise}
\label{subsec:compromise}

We adapt the mechanism above to simulate $\*Z_{1}, \dots, \*Z_{B}$ in the case where we have more than one trained feature extractor $T\left(\cdot, \hat{\theta}_s\right)$, for $s = 1, \dots, S$. Set $S < B$, so the feature learning phase is less costly than in section \ref{subsec:nonparametric_bootstrap}.

Begin by extracting $\*Z_{s}$ on resampled versions of the inference set, using the $S$ extractors $T\left(\cdot, \hat{\theta}_{s}\right)$. Then compute their truncated, rank-$K$ SVDs $\hat{\*U}_{s}\hat{\*\Sigma}_{s}\*V_{s}^T$. New feature sets are simulated from,
\begin{align*}
\*Z_{b}= \left(\hat{\*U}_{s\left(b\right)}\hat{\*\Sigma}_{s\left(b\right)} + \*E_{b}\right)\*\Pi_{b},
\end{align*} 
where $s\left(b\right)$ is drawn uniformly from $1, \dots, S$ and  $\*E_{b}$ resamples entries across all $\*Z_{s} - \hat{\*Z}_{s}$.
Given the $B$ resampled $\*Z_{b}$, we generate confidence regions as before.

\section{Simulations}
\label{sec:simulation}

We conduct two simulation studies. The first uses a low-rank model and permits calculation of coverage rates, but it is less representative of realistic feature learning settings. The second generates images using a spatial point process with variation reflecting a small set of latent parameters. The distributed feature learning associated with this setting prevents us from computing the coverage of confidence ellipsoids, but its complexity more accurately reflects practice.

\subsection{Low-rank model}
\label{subsec:low_rank_model}

The first simulation generates samples $\*X \in \reals^{n \times D}$ using,
\begin{align}
\label{eq:low_rank1}
  \*X &= \*U\*\Sigma \*V^{T} + \*E & \*\Sigma &= \text{diag}\left(c\*1_{K}\right) & \*U &\sim \Haar\left(n, K\right)\\
  y & = \*U \*\Sigma \beta + \epsilon & \beta &= \left(b \*1_{\frac{K}{2}}, -b \*1_{\frac{K}{2}}\right) & \*V &\sim \Haar\left(D, K\right)\\
  &&&& E_{ij} &\sim \Gsn\left(0, \sigma_{\*E}^{2} \right)\\
  &&&& \epsilon_{i} &\sim \Gsn\left(0, \sigma_{y}^2\right)
\end{align}
where $\Haar\left(n, K\right)$ denotes a random orthonormal matrix with $n$ rows and $K$ columns. $\*X$ is a random rank-$K$ matrix observed with Gaussian noise. 
$y$ is a response depending on the latent coordinates of each row. The specific parameters we use are $N = 1000, D = 100, c = 100, b = 1, K = 2, \sigma^2_{E} = 0.1$. Note that this is a model of the data $\*X$, not the features $\*Z$, as in equation \ref{eq:para_boot}.

As a feature extractor, we use a randomly perturbed and permuted SVD-based estimate of the latent coordinates,
\begin{align}
\begin{split}
\label{eq:low_rank2}
\*X &= \hat{\*U}\hat{\*\Sigma}\hat{\*V}^T \\
\*Z &:= \left(\hat{\*U}_{\hat{K}}\hat{\*\Sigma}_{\hat{K}} + \tilde{\*E}\right) \*\Pi
\end{split}
\end{align}
where the subscript $\hat{K}$ denotes that only the top $\hat{K}$ left singular vectors and values are used and $\tilde{E}_{ij} \sim \Gsn\left(0, 0.1^2\right)$. The permutation $\*\Pi$ and noise $\tilde{\*E}$ mimic the variation across retrained feature extractors. Given this source data and feature extractors, we apply all three bootstrap methods to this data, generating $B = 1000$ bootstrap replicates in each case. For the compromise approach, we train $S = 100$ separate feature extractors.

\begin{figure}
    \centering
    \includegraphics[width=\textwidth]{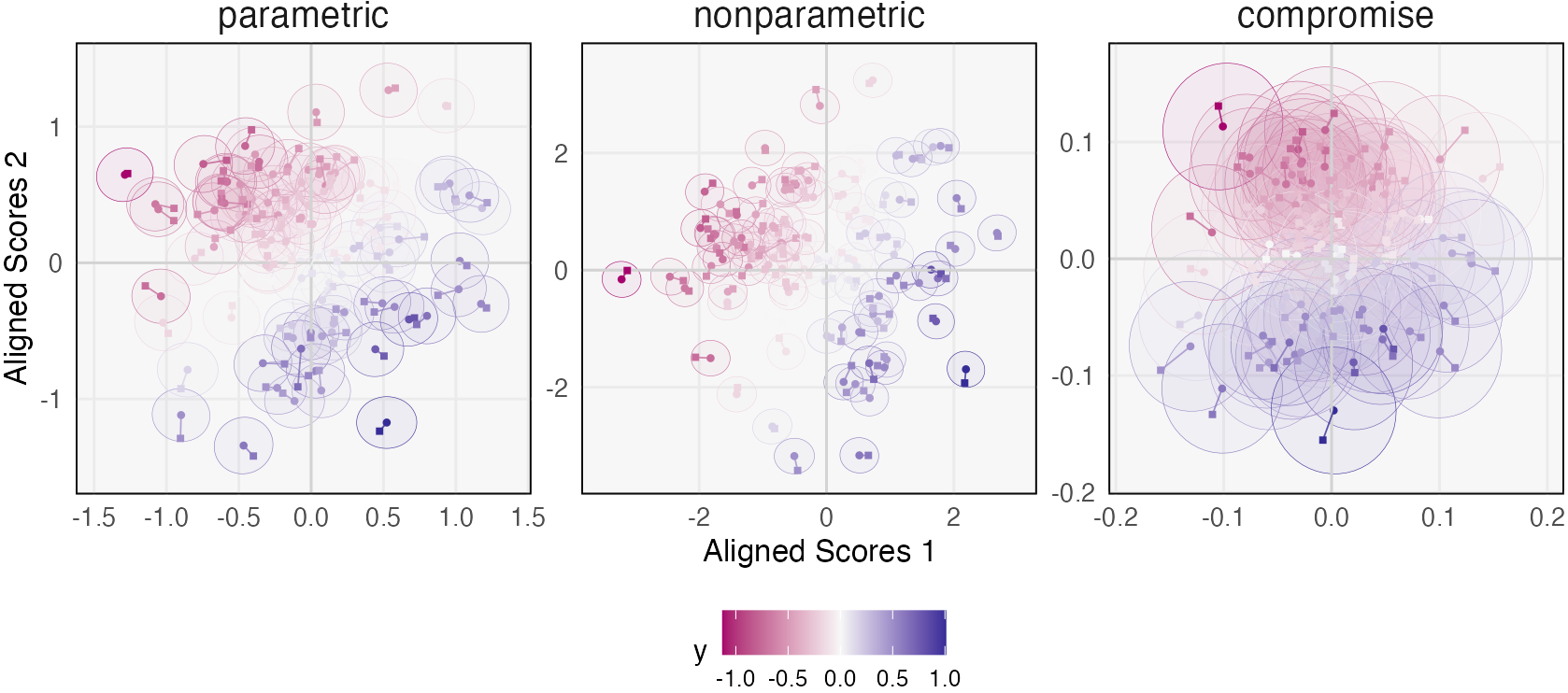}
    \caption{Projections from the low-rank data simulation. Each ellipse gives the confidence area for the latent coordinates of one sample. Squares are the positions of the true low-rank coordinates after Procrustes rotating them to align with the centers of the ellipses. The confidence areas for the nonparametric bootstrap are smaller than those for the parametric bootstrap. Those for the compromise method are conservative.}
    \label{fig:low_rank_projections}
\end{figure}

The resulting 95\% confidence ellipses are shown in Figure \ref{fig:low_rank_projections}. Qualitatively, the parametric and nonparametric bootstrap approaches provide similar output. A gradient across the colors of $y$ reflects accurate estimation of the true latent factors. We have Procrustes aligned the true coordinates $\*U \*\Sigma$ (squares) with the $B$ bootstrap replicates, and the fact that most squares are contained within ellipses suggests that the bootstrap accurately reflects uncertainty in the estimated projections. In fact, for the parametric and nonparametric approaches, the empirical coverage rates of these ellipses are 96.4\% and 95.2\%, respectively. On the other hand, the compromise approach appears to be overly conservative, with a coverage of 99.9\%. This behavior arises in the remaining simulations and data analysis as well.

\subsection{Spatial point process}
\label{subsec:point_process}

\begin{figure}
  \centering
  \includegraphics[width=\textwidth]{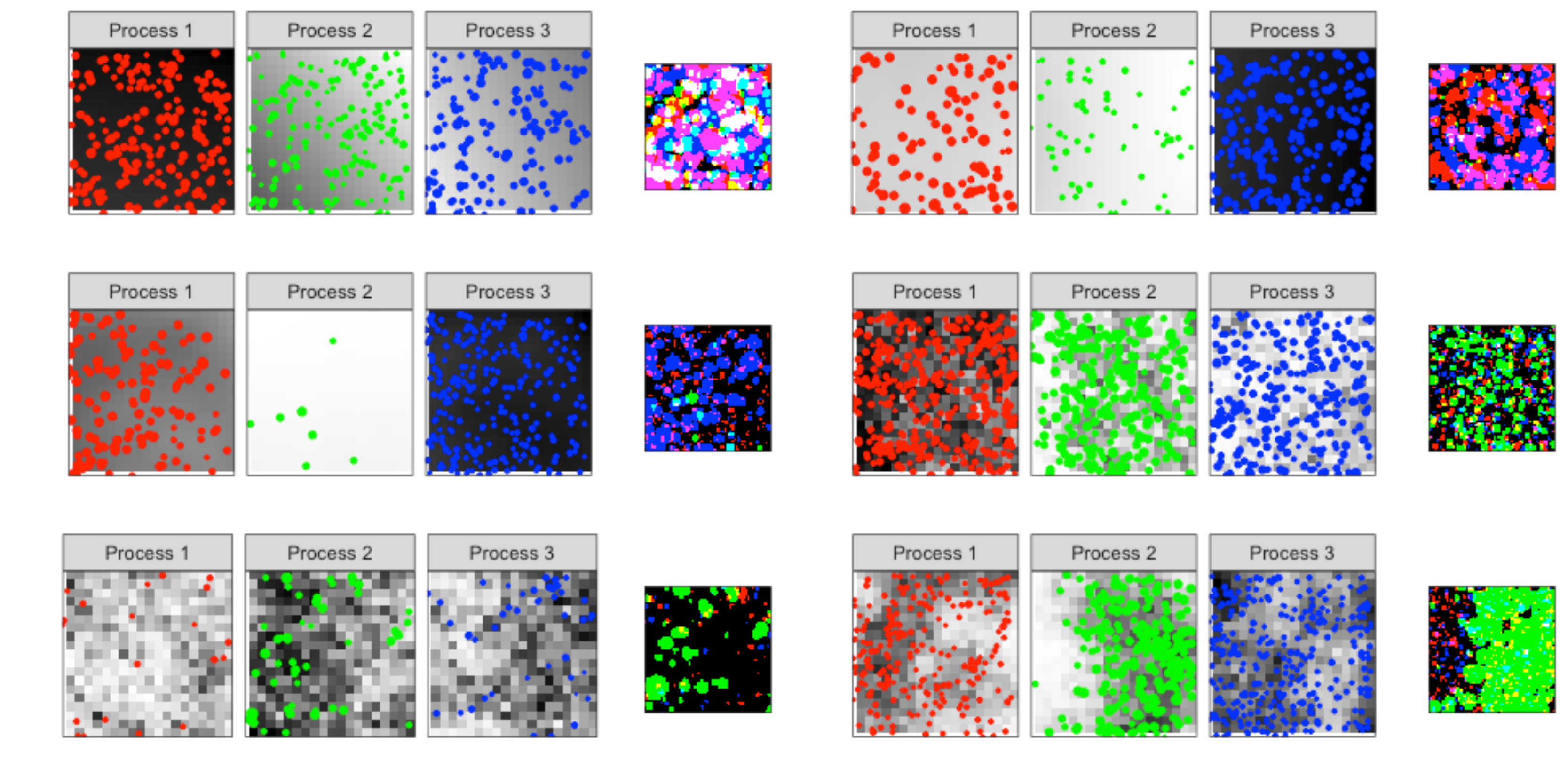}
  \caption{Example images, for low (top), average (middle), and high
    (bottom) values of $y_i$. For each sample, three relative intensity
    functions $B_{r}\left(x\right)$ are generated, shown as a greyscale
    heatmaps. Samples drawn from each process are overlaid as circles. The final
    images combine points across
    processes, removing the underlying intensity function, which is not
    available to the feature learner. Small $y_i$ values are associated with
    smoother, less structured intensity functions.}
  \label{fig:matern_example}
\end{figure}

In this simulation, we generate a collection of images using a point process where parameters vary from image to image. Intuitively, each image represents cells viewed through a microscope, and different latent parameters influence the cell ecosystem. A single response value $y$ is associated with these latent parameters. Example images for varying $y$ are given in Figure \ref{fig:matern_example}. We generate 10,000 of these $64 \times 64 \times 3$-dimensional RGB images.

\subsubsection{Generation}
\label{subsubsec:generation}

Locations of cells are governed by an intensity
function drawn from a two-dimensional marked Log Cox Matern Process (LCMP)
\cite{diggle2013spatial}. Recall that a Matern process is a Gaussian process
with covariance function,
\begin{align}
  \label{eq:cov_lcmp}
C_{\nu, \alpha}(\|x - y\|)=\sigma^{2} \frac{2^{1-\nu}}{\Gamma(\nu)}\left(\sqrt{2 \nu} \frac{\|x - y\|}{\alpha}\right)^{\nu} K_{\nu}\left(\sqrt{2 \nu} \frac{\|x - y\|}{\alpha}\right),
\end{align}
where $\alpha$ acts like a bandwidth parameter and $\nu$ controls roughness.

Our LCMP has $R$ classes (cell types). This can be constructed as follows. First, a nonnegative process $\Lambda\left(x\right)$ is simulated along the image grid,
$\Lambda\left(x\right) \sim \exp{\mathcal{N}\left(0, \mathbf{C}_{\nu_{\Lambda},
    \alpha_{\Lambda}}\right)}$, where $\mathbf{C}_{\nu_{\Lambda},
  \alpha_{\Lambda}}$ is the covariance matrix induced by equation \ref{eq:cov_lcmp}. This is a baseline intensity
that determines the location of cells, regardless of cell type. $R$
further processes are then simulated, $B_{r}\left(x\right) \sim \exp{\beta_{r} +
  \mathcal{N}\left(0, \mathbf{C}_{\nu_{B}, \alpha_{B}}\right)} $. These
processes reflect relative frequencies of the $R$ classes at any
location $x$; the intercept $\beta_r$ makes a class either more or less frequent
across all positions $x$. Given these intensity functions, we can simulate $N$ cell locations by drawing
from an inhomogeneous Poisson process with intensity $\Lambda\left(x\right)$.
For a cell at location $x$, we assign it cell type $r$ with probability
$\frac{B_{r}^{\tau}\left(x\right)}{\sum_{r^\prime = 1}^{R}
  B^{\tau}_{r^\prime}\left(x\right)}$. We have introduced a temperature
 $\tau$ controlling the degree of mixedness between cell types at a
given location.

To complete the procedure for simulating images, we add two last sources of
variation — the number of cells and cell size. The number of cells per image
is drawn uniformly from 50 to 1000. The cells from class $R$ are drawn with a
random $\text{Gamma}\left(5, \lambda_{r}\right)$ radius. A summary of
all parameters used to generate each image is given in Supplementary Table
\ref{tab:sim_params}. Each parameter is drawn uniformly within its range, which
has been chosen to provide sufficient variation in image appearance. These
parameters are the ``true'' underlying features associated with the simulated
images; they give the most concise description of the variation observed across
the images. The response $y$ is a hand-specified linear combination of these parameters, the reasoning behind the combination is discussed in the \texttt{generate.Rmd} script in the accompanying compendium, see Supplementary Section \ref{sec:reproducibility}.

\subsubsection{Experiments}
\label{subsubsec:experiments}

We study the influence of the following parameters,
\begin{itemize}
\item Learning vs. inference split sizes. We vary the proportion of data used for learning and inference. We sample $I$ so that $\frac{1}{n}\absarg{I} \in \{0.15, 0.5, 0.9\}$.
\item Models trained. For feature extractors, we train CNN, VAE, and RCF models on the learning split $I$.
\item Model complexity. We train VAEs whose hidden layer has dimensionality $L \in \{32, 64, 128\}$. Similarly, we vary the number of first-layer convolutional filters in the CNN model across $L \in \{ 32, 64, 128\}$. For the RCF, we use $L \in \{256, 512, 1024\}$ random features. This increase reflects the fact that more random features must be considered before a subset of predictive ones are identified.
\item Inference strategy. We use the parametric, nonparametric, and compromise bootstrap strategies from section \ref{sec:methods} to estimate confidence areas for the projections obtained by feature learners.
\end{itemize}

\begin{figure}
  \centering
  \includegraphics[width=\textwidth]{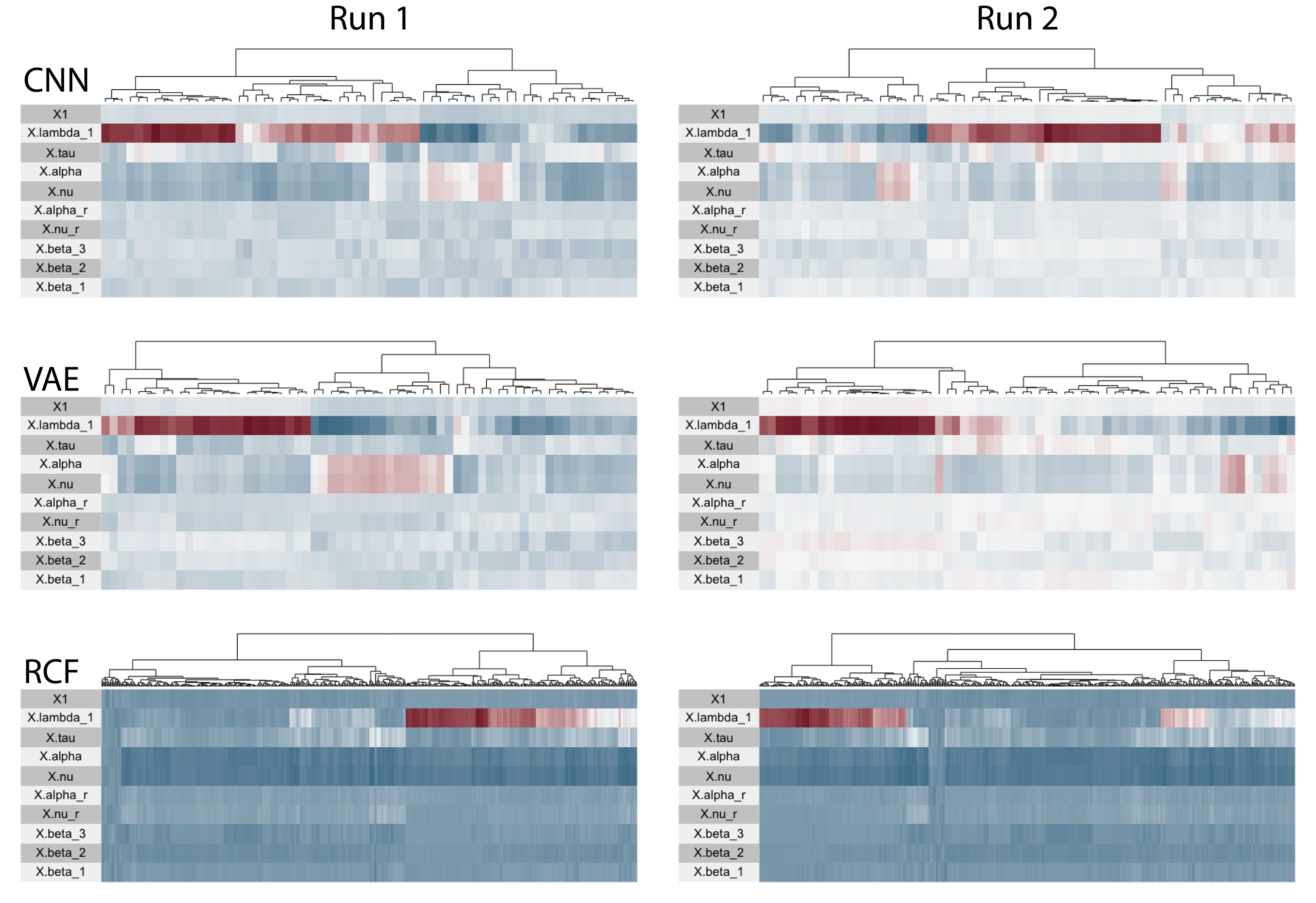}
  \caption{Each feature learning algorithm learns a distributed representation of the true underlying features in the simulation. Within each heatmap, rows correspond to the parameters in Supplementary Table \ref{tab:sim_params}. Columns are activations of learned features; they have been reordered using the package \protect\cite{barter2018superheat}. The color of a cell gives the correlation between true and learned features. Blue and Burgundy encode positive and negative correlation, respectively.} 
  \label{fig:distributed_hm}
\end{figure}

Figure
\ref{fig:distributed_hm} shows the activations of learned features across 2000
images for two perturbed versions of the training data when 90\% of the data are used for inference and $L = 64$ (CNN, VAE) and $512$ (RCF). The learned features correspond to,
\begin{itemize}
\item CNN: Activations from the final hidden layer of neurons, used directly as input
  for the regression.
\item VAE: Spatially-pooled activations from the middle, encoding layer of the
  variational autoencoder.
\item RCF: The spatially-pooled activations corresponding to each random
  convolutional feature.
\end{itemize}

Note that, across algorithms, there is no simple correspondence between learned and source features (i.e., parameters of the
underlying simulation). Instead, there are clusters of learned features,  corresponding to a pattern across multiple source features. We also
find subsets of features across all models that are only weakly
correlated with any source feature. This has been referred to as distributed representation learning \cite{hinton1984distributed, le2014distributed}.

Certain source features appear ``easier'' to represent than others,
in the sense that more of the learned features are strongly correlated with
them. Many features are correlated with $N_{i}$, the total number of cells in
the image, and $\lambda_{i1}$, the size of the cells from Process 1. Depending
on the model, the bandwidth $\alpha_{ir}$, roughness $\nu_{ir}$, and prevalence
$\beta_{ik}$ parameters are either only weakly or not at all correlated with
learned features. Even when
features detect variation in $\alpha_{ir}$ and $\nu_{ir}$, they cannot
disambiguate between these two parameters.
Finally, the CNN and
VAE features tend to be more clustered, with strong correlation across several
source features. In contrast, the RCF features show more gradual shifts in
correlation strength. They also show relatively little variation in correlation
strength across features other than $\lambda_{i1}$ and $N_{i}$.

\begin{figure}
    \centering
    \includegraphics[width=\textwidth]{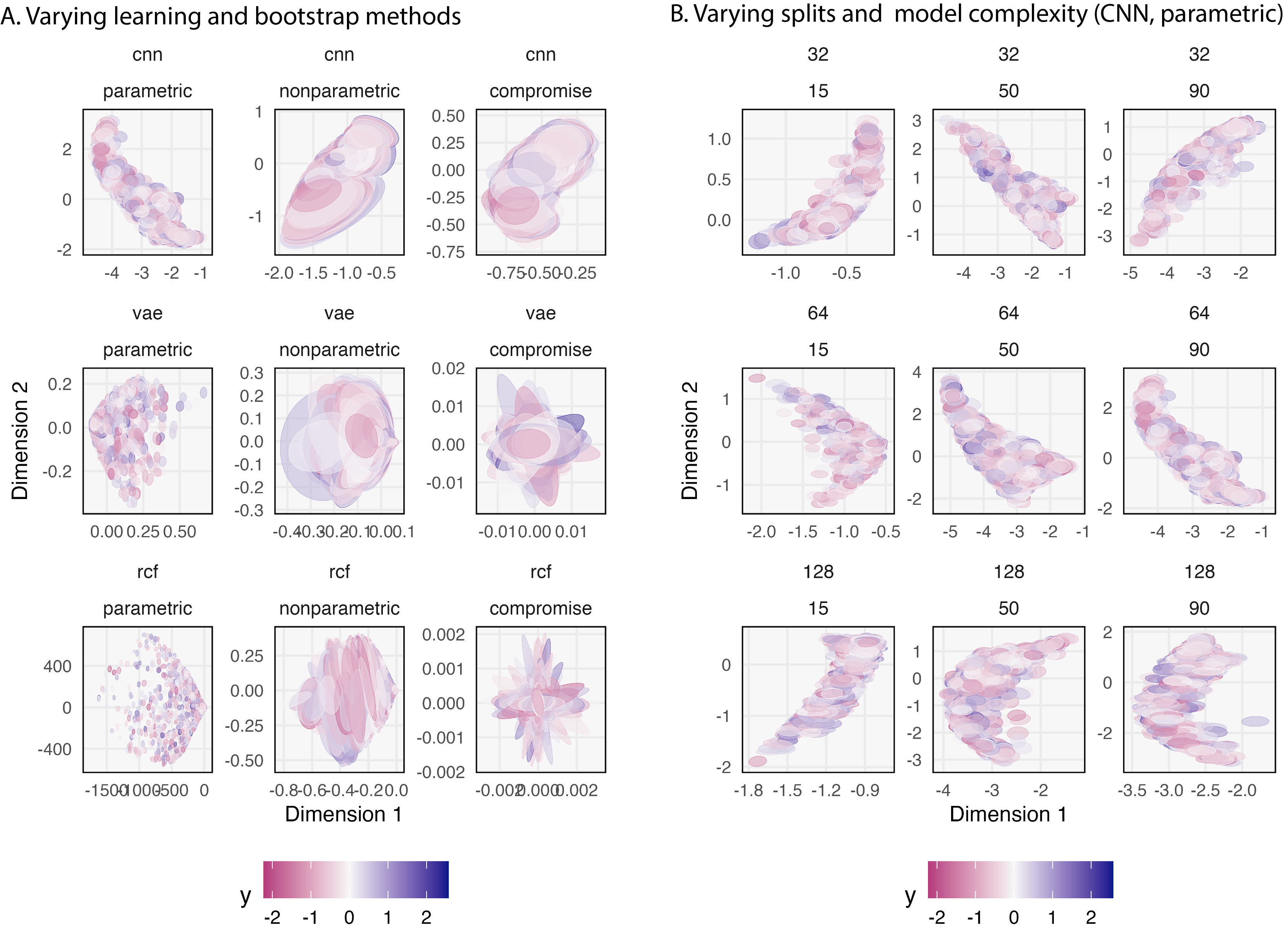}
    \caption{(a) The 95\% confidence areas associated with projections from the spatial point process simulation. Each point correspond to one image. Only the setting with 90\% of the data used for feature learning and the midsize models ($L = 64$ for the CNN and VAE, $L = 512$ for the RCF) are shown. (b) A view of confidence areas for the CNN across a range of learning split fractions ($0.15, 0.5, 0.9$) and model complexities ($L = 32, 64, 128$), all using the nonparametric approach.}
    \label{fig:simulation_projection_combined}
\end{figure}

Example confidence areas across models and bootstrapping approaches are given in Figure \ref{fig:simulation_projection_combined}a. In contrast to the Figure \ref{fig:low_rank_projections} of the low-rank simulation, the areas from the nonparametric bootstrap are larger than those from the parametric bootstrap. 
This disagreement suggests that the proposed mechanism of equations \ref{eq:low_rank1} and \ref{eq:low_rank2} is insufficient for characterizing differences in learned features that arise between runs of more complex feature extractors -- multiple runs must be used for to account for randomness in algorithmic feature learning. As before, the compromise bootstrap has larger confidence areas than either parametric or nonparametric approaches on their own. In general, the RCF tends to have smaller confidence areas compared to the CNN and VAE.

Figure \ref{fig:simulation_projection_combined}b shows confidence regions for a single model (CNN) and bootstrap procedure (parametric) across a range of model complexities and split proportions. For larger $L$, projections are further from the origin, suggesting larger activations on average. The fraction of data used for feature learning does not appear to affect the strength of the association with the response $y$ or the size of the projection uncertainties. Corresponding figures for the other models are provided in Supplementary Section \ref{sec:supplementary_figures}.

\section{Data Analysis}
\label{sec:data_analysis}

We next analyze the spatial
proteomics dataset reported in 
\cite{keren2018structured}, which found a relationship between the spatial
organization of Triple Negative Breast Cancer (TNBC) tissues and disease
progression. In a classical proteomics study, the expression levels for a set of
proteins is measured for a collection of cells, but the cell locations are
unknown. In contrast, these data provide for each patient (1) an image
delineating cell boundaries and (2) the protein expression levels associated
with each cell in the images.

We only work with spatial cell delineations, not protein expression
levels. This allows us to study feature learning within the
images without having to worry about linking expression and image data,
which is itself a complex integration problem. The data are $2048 \times
2048$-dimensional images, one for each of 41 patients. Each pixel has a
value 1 through 7 encoding which of 7 categories of tumor or immune cell
types the pixel belongs. To ensure that the cell types are treated as
categorical, we transform pixels to their one-hot encodings, resulting in a
collection of $2048 \times 2048 \times 7$ binary matrices.
\begin{figure}
  \centering
  \includegraphics[width=\textwidth]{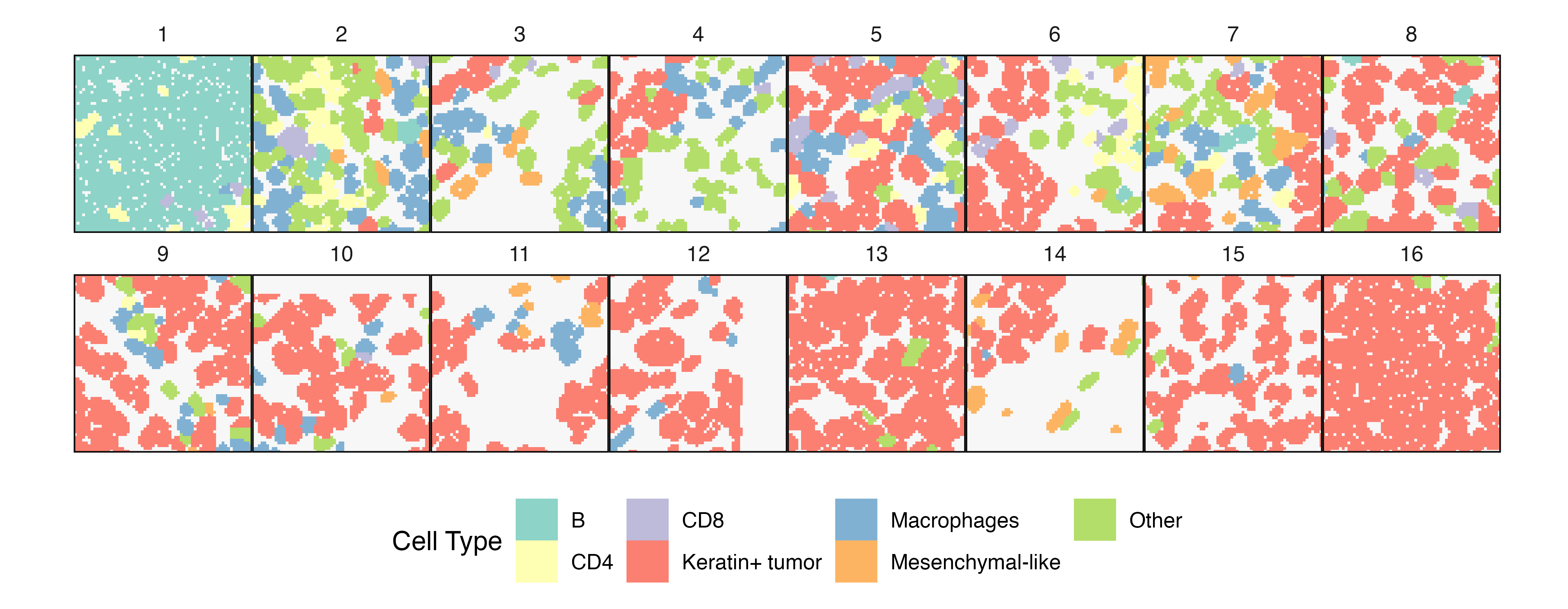}
  \caption{Example patches from the TNBC data. Panels are ordered by $y_i$, the
    (log) fraction of cells they contain that belong to the tumor. This provides signal for the supervised algorithms, whose goal is to correctly place patches from
    new patients along this gradient.}
  \label{fig:example_cells}
\end{figure}

To setup a prediction problem, we split each image into $512 \times 512 \times 7$
patches. These patches are our $x_{i}$. Patches from 32 of the patients are
reserved for feature learning. Four among these 32 are used as a development
split, to tune parameters of the feature learning algorithms. As a response
variable, we use $y_{i} = \log\left(\frac{\#\{\text{Tumor cells in
}x_{i}\}}{\#\{\text{Immune cells in }x_i\}}\right)$. These $y_i$ provide
signal for the supervised feature learners. Example cell patches
are shown in Figure \ref{fig:example_cells}. We fit the same models (CNN, VAE, and RCF) as discussed in section \ref{sec:simulation}, varying model complexity over the same parameters as before.

\begin{figure}
  \centering
  \includegraphics[width=\textwidth]{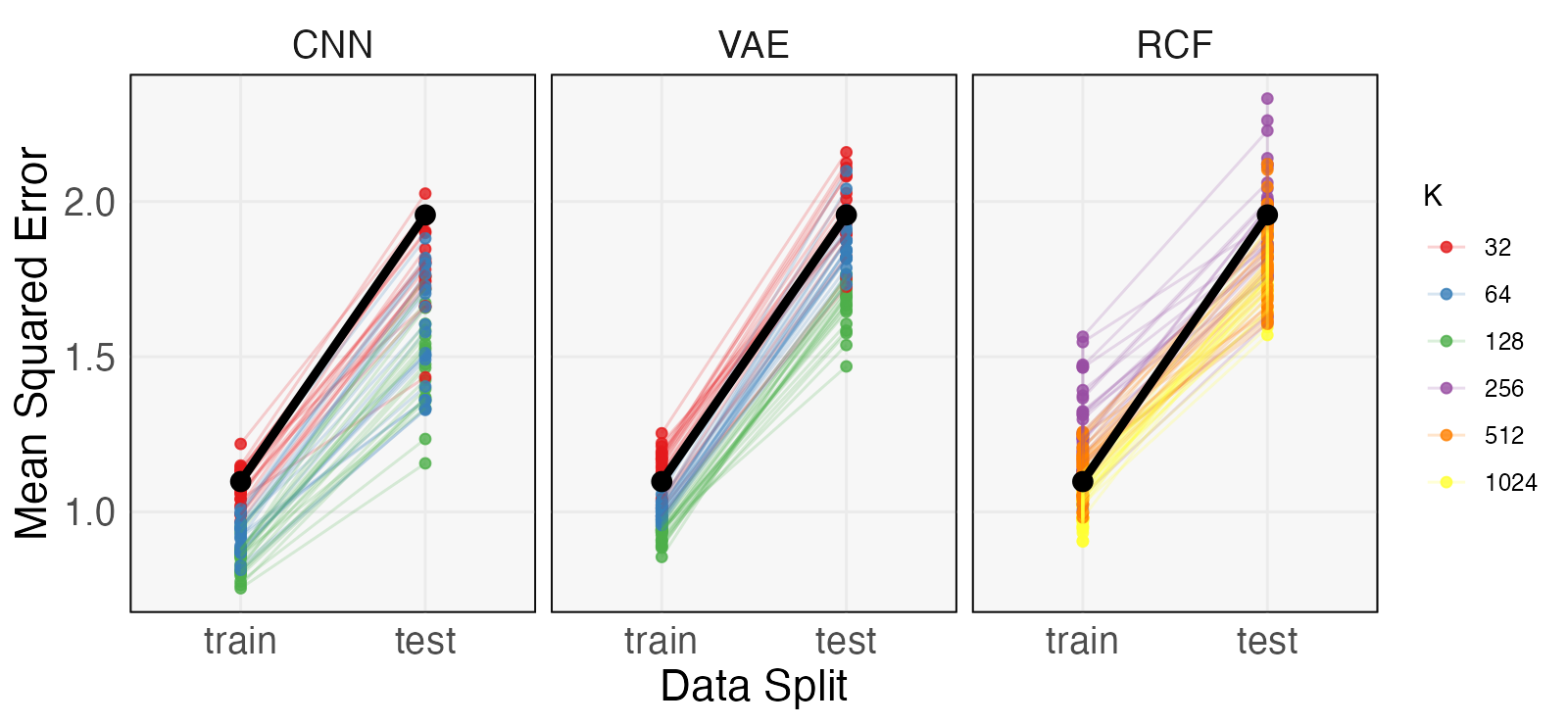}
  \caption{Relative performance of feature learning strategies on the TNBC data.
    Linked points come from the same bootstrap replicate. The black line gives the baseline ridge regression approach using the manually generated raw pixel counts for
    each of the cell types as predictors. Predictions from the development split are omitted;
    this split has few patches, and the estimated MSE's have high variance.}
  \label{fig:tnbc_baseline}
\end{figure}

As a baseline, we compare against a ridge regression with pixelwise composition
features. We train a model with $y$ as a response and the
average number of pixels belonging to each of the cell-type categories as a
$7$-dimensional feature vector. This helps to determine whether the model has
learned interesting features for counting cells, like cell size and
boundaries, rather than simply averaging across pixel values. Indeed, Figure
\ref{fig:tnbc_baseline} shows that, except in models with low capacity $L$, performance is improved when learning features algorithmically.

\begin{figure}
    \centering
    \includegraphics[width=\textwidth]{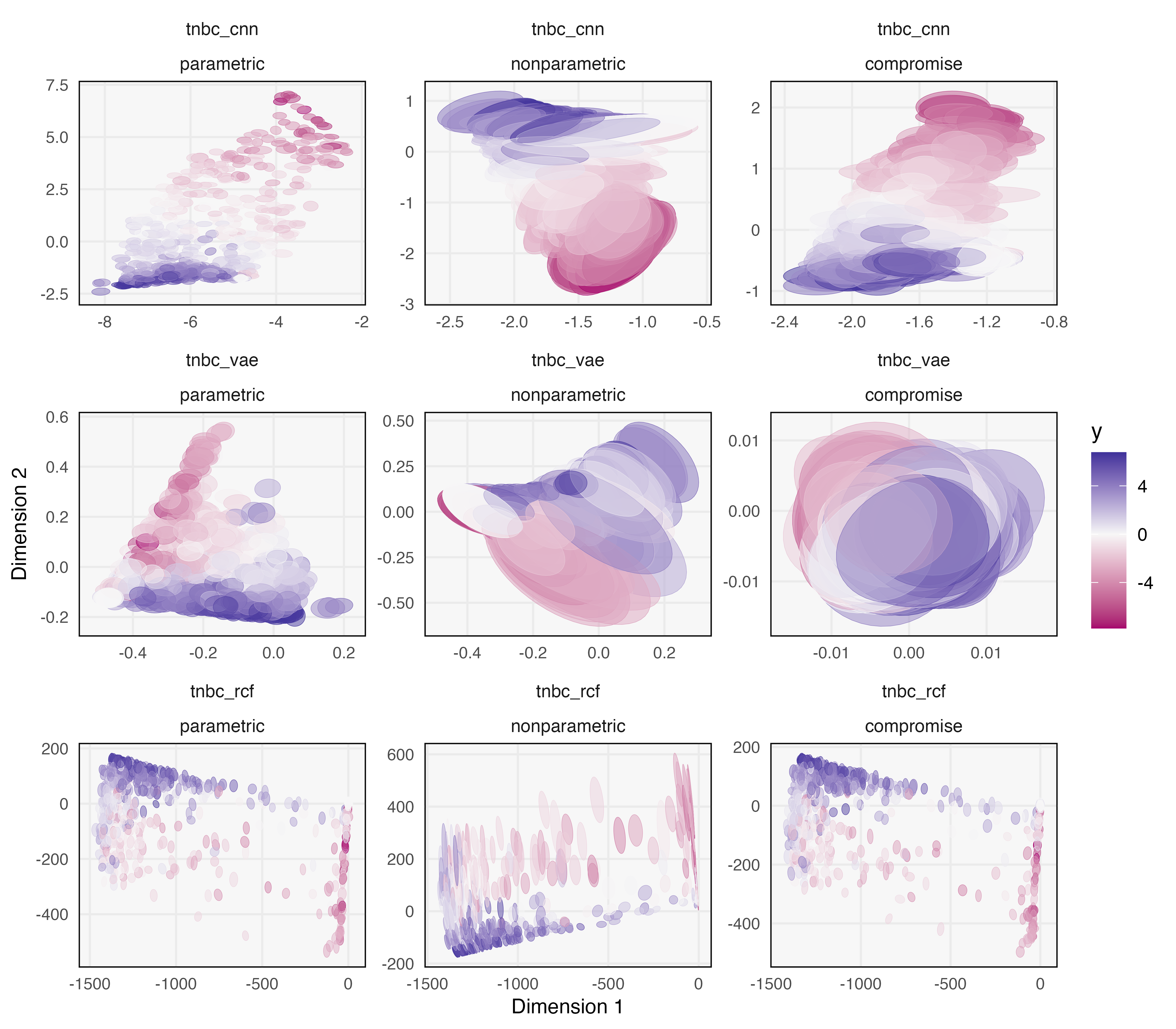}
    \caption{Confidence areas from the TNBC application. Points are shaded by $y_{i} = \log\left(\frac{\#\{\text{Tumor cells in
}x_{i}\}}{\#\{\text{Immune cells in }x_i\}}\right)$ , which provides the supervisory signal to the CNN and RCF during feature extraction. Models and bootstrap procedures are arranged along rows and columns, respectively. Only the models with intermediate complexity ($L = 64$ for the CNN and VAE, $L = 512$ for the RCF) are shown. Analogous figures for other $L$ are given in the supplementary materials.}
    \label{fig:64_coordinates}
\end{figure}

To characterize features and their uncertainty, we perform $B = 100$ iterations for each of the parametric, nonparametric, and compromise bootstrap strategies. In each case, the samples are generated from patches reserved for inference. Two-dimensional projections for fixed model complexities are given in Figure \ref{fig:64_coordinates}. As visible by the color gradients, all methods learn to differentiate between patches with small and large values of $y_{i}$, even the VAE, which is unsupervised. Comparing rows, the RCF appears to give the most stable representations, while the coordinates for the VAE have larger confidence areas in general. For all models, some projections appear to be more uncertain than others. Moreover, the certain axes directions tend to be more uncertain than others, reflected by the eccentricity of ellipses. For example, viewing estimates from the nonparametric approach, the VAE projections have the highest uncertainty for low values of Dimension 2. Analogously, high values of Dimension 2 have high uncertainty in the RCF. 

For the CNN and RCF, the three bootstrap approaches give qualitatively similar conclusions about regions with higher and lower uncertainty, though the average sizes of the confidence areas differ. The size of confidence areas for the compromise approach in this case seem intermediate between those of the parametric and nonparametric approaches. For the VAE, the bootstrap approaches do not appear to agree. The compromise approach generally gives much larger confidence areas, potentially reflecting a failure of the Procrustes alignment in this case. Though this figure only displays one $L$ for each model, we find few differences in the projection uncertainty across models with different complexities, see Supplementary Figure \ref{fig:nonparametric_coordinates} in the supplementary materials.

Figure \ref{fig:tnbc_imagegrid-PCA-1-2} overlays example patches onto aligned
coordinates. In the CNN, samples in the bottom right have a high fraction of immune cells,
those on the top left are mostly tumor, and those at the top right have lower cell density. In light of the confidence regions in Figure \ref{fig:nonparametric_coordinates}, embeddings of immune cell rich images tend to show more variability across bootstraps. In the RCF, the first dimension similarly reflects the tumor vs. immune gradient. The lower uncertainty of projections along this axis suggests that this gradient is reliably captured across runs of the feature extractor.
The lower right region of the VAE has high cell diversity, and the upper left has lower density. It seems that regions with higher cell diversity and larger proportions of immune cells also have more uncertain embeddings. From the top right to the bottom left, there is again an tumor to immune transition.

\begin{figure}
  \centering
  \makebox[\textwidth][c]{\includegraphics[width=\textwidth]{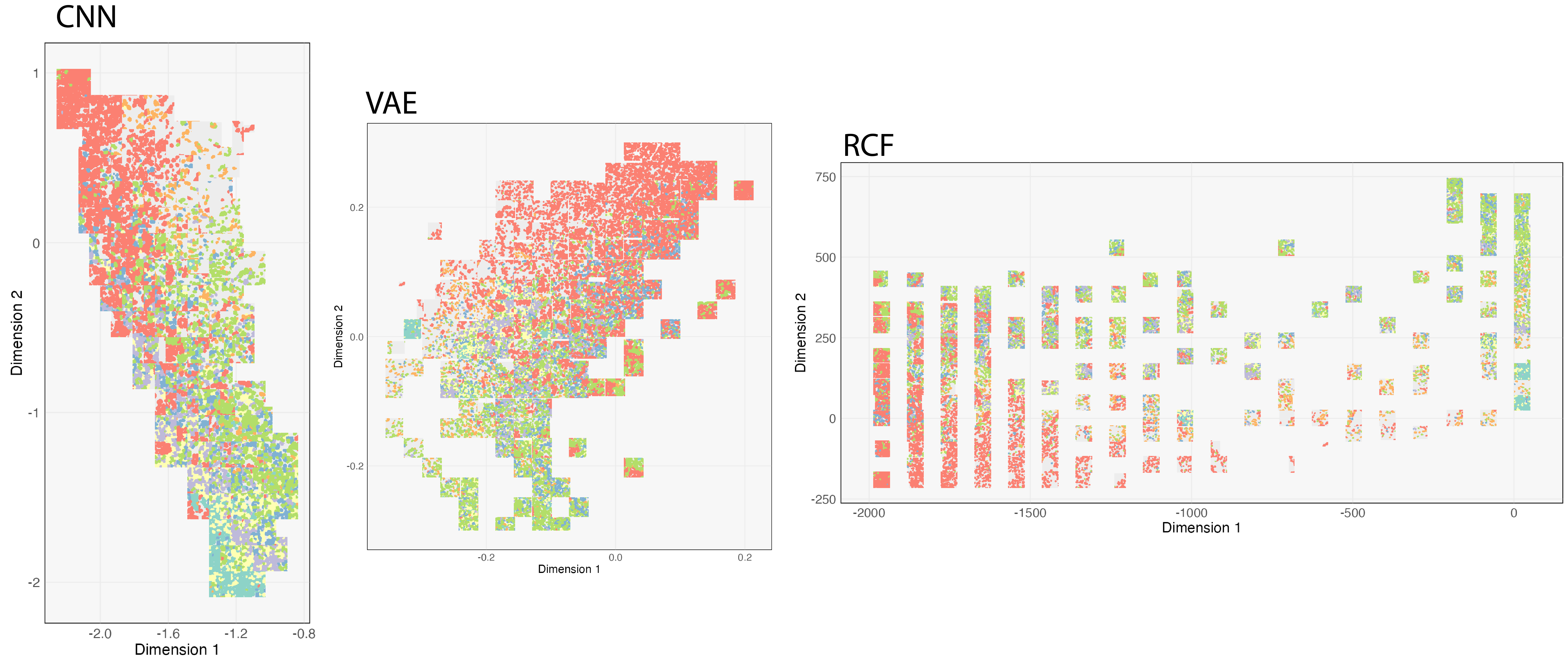}}
  \caption{A version of the nonparametric bootstrap column from the dimensionality reductions in Figure \ref{fig:64_coordinates},
    overlaying representative samples across the learned feature space. Cells
    are color coded as in Figure \ref{fig:example_cells}. Note that the overall
    shape of the region in which images are displayed mirrors the shape of the
    cloud of points in Figure \ref{fig:64_coordinates}.}
  \label{fig:tnbc_imagegrid-PCA-1-2}
\end{figure}

\section{Discussion}
\label{sec:discussion}

We have adapted existing approaches for evaluating the uncertainty of projections in dimensionality reduction for use in the context of algorithmically learned features. We have conducted an empirical study using simulations of varying complexity and a spatial proteomics data analysis problem, applying a representative suite of feature learning algorithms. We found that in more complex settings, a parametric bootstrap based on a single set of learned features does not reflect the degree of uncertainty present when comparing features derived from independently trained models.

Our results raise several questions for further study. It is natural to ask
to what extent similar behaviors are exhibited across other data domains, model
types, or training regimes. For example, it would not be unusual to represent
the cell data in our case study using a marked graph linking neighboring cells.
Do the features learned by a graph autoencoder have similar stability
properties? More generally, are there classes of feature learning algorithms that all share behavior from the stability perspective? 
In other domains, we may ask whether our methods can be adapted to text or audio data.

Though the proposed bootstraps provide similar conclusions in the low-rank simulation, they differ in the point process simulation and spatial proteomics data analysis. This suggests that mechanisms in equations \ref{eq:low_rank1} and \ref{eq:low_rank2} may not reflect the behavior of repeated feature learning in more complex situations. 
The assumption that features can be rotated to align runs may also be problematic, and more general transformations across feature learning runs are plausible. For the CNN and RCF models in the data analysis example, we find that confidence areas for the compromise approach are intermediate between those for the parametric and nonparametric bootstraps. 
However, in both simulations, it tends to be larger, and we have no explanation. Finally, though we have empirically found coverage rates for the parametric and nonparametric bootstraps to be acceptable in the low-rank simulation, we have not theoretically studied the properties of these procedures. This could clarify differences that arise in projection uncertainties between bootstrap and feature learning approaches.

\section*{Acknowledgments}

The author thanks Susan Holmes, Karl Rohe, three reviewers, and the editor for feedback which improved the manuscript. Research was performed with assistance of the UW-Madison Center For High Throughput Computing (CHTC).

\bibliographystyle{apacite}
\bibliography{refs}

\section{Reproducibility}
\label{sec:reproducibility}

We have prepared a research compendium at \href{https://github.com/krisrs1128/LFBCR}{https://github.com/krisrs1128/LFBCR}.
Instructions for downloading raw and preprocessed data are given in the \texttt{analysis/data/} subdirectory.
The \texttt{LFBCR} compendium available in the repository above is also an R package and includes functions for that can be used to apply the bootstrap and alignment methods more generally.

Rmarkdown files needed to reproduce the figures in Section \ref{sec:simulation} are given in the \texttt{analysis/simulations} subfolder. The analogous files for the spatial proteomics application are given in \texttt{analysis/data\_analysis}. The \texttt{learning} subfolders at both these locations
can be used to retrain one instance of each of the feature learning algorithms. Further details are
given in the \texttt{README.md} folders in the repository.

\section{Supplementary Tables and Figures}
\label{sec:supplementary_figures}

\begin{table}[]
\begin{tabular}{|p{0.1\linewidth}|p{0.4\linewidth}|p{0.16\linewidth}|p{0.15\linewidth}|}
\hline
\textbf{Feature}              & \textbf{Description}                                 & \textbf{Influence}          & \textbf{Range}             \\
\hline
$N_i$                & The total number of cells.                                    & 0.5                         & $\left[50, 1000\right]$    \\
\hline
$\nu_{i,\Lambda}$    & The roughness of the overall intensity process.            & -0.5                        & $\left[0, 8\right]$        \\
\hline
$\alpha_{i,\Lambda}$ & The bandwidth of the overall intensity process.            & -0.5                        & $\left[0, 8\right]$        \\
\hline
$\beta_{ir}$             & The intercept controlling the frequency of class $r$. & 1 for $r = 1$,\ -1 otherwise & $\left[-0.15, 0.15\right]$ \\
\hline
$\nu_{iB}$           & The roughness of the relative intensity processes.            & -0.5                        & $\left[0, 3\right]$        \\
\hline
$\alpha_{iB}$        & The bandwidth of relative intensity processes.                & -0.5                        & $\left[0, 3\right]$        \\
\hline
$\tau_{i}$           & The temperature used in cell type assignment.                 & 0.5                         & $\left[0, 3\right]$        \\
\hline
$\lambda_{ir}$       & The shape parameter controlling the sizes of each cell type.  & 1 for $r = 1$,\ 0 otherwise  & $\left[100, 500\right]$   \\
\hline
\end{tabular}
\caption{Our simulation mechanism is governed by the above parameters.
  Parameters $N_i$ and $\lambda_{ir}$ control the number and sizes of imaginary
  cells. $\nu_{i, \Lambda}$, $\alpha_{i, \Lambda}$, $\beta_{ik}$, $\nu_{iB}$,
  $\alpha_{iB}$, and $\tau_{i}$ control the overall and relative intensities of
  the marked LCMP from which the cell positions are drawn. Example draws are
  displayed in Figure \ref{fig:matern_example}.}
\label{tab:sim_params}
\end{table}

\begin{figure}
    \centering
    \includegraphics[width=\textwidth]{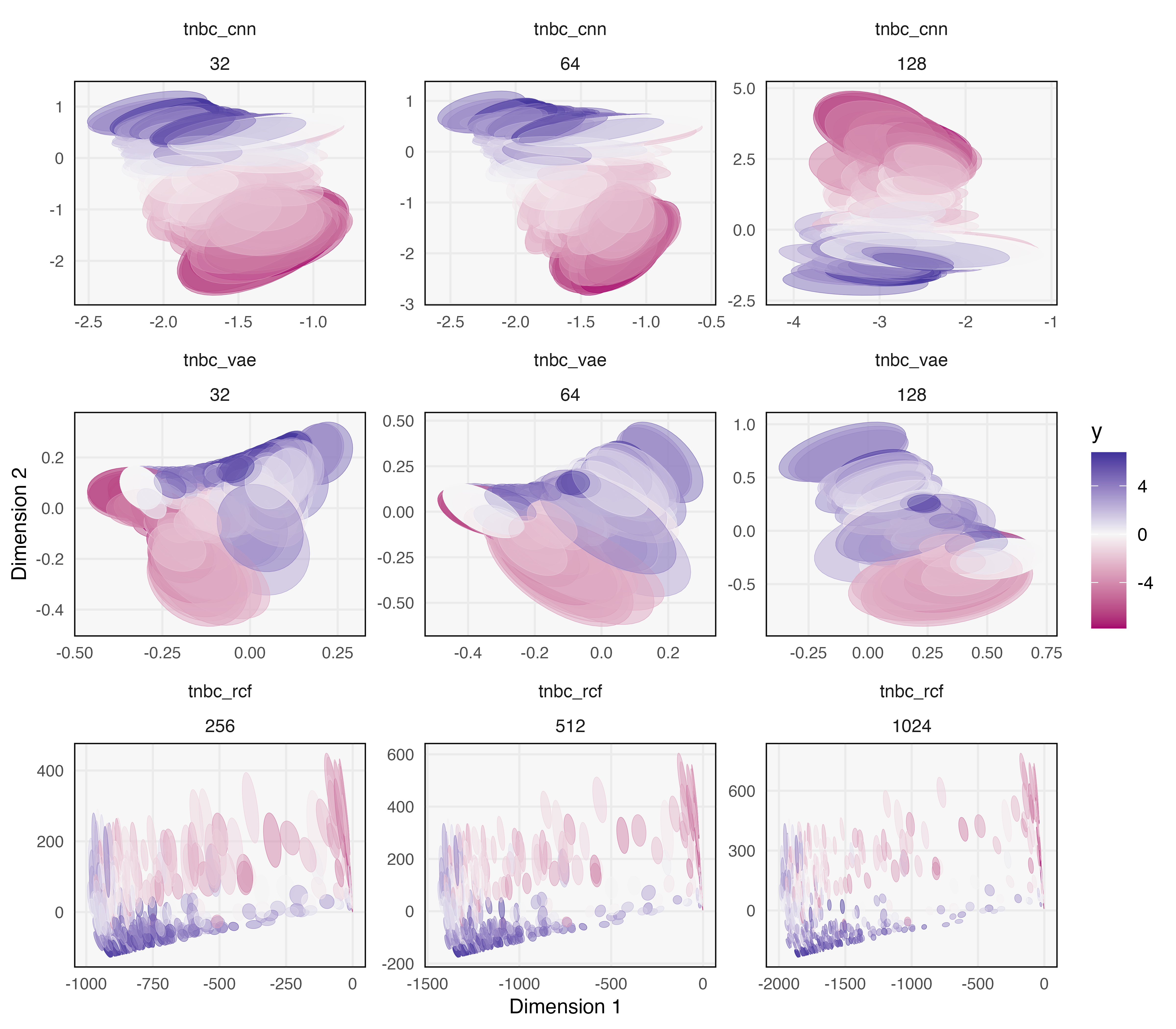}
    \caption{A comparison of the influence of models (rows) and model complexities (columns) on confidence regions constructed using the nonparametric bootstrap in the data analysis application. Each individual panel is read similarly to Figure \ref{fig:64_coordinates}.}
    \label{fig:nonparametric_coordinates}
\end{figure}

\begin{figure}
    \centering
    \subfloat{\includegraphics[width=0.5\textwidth]{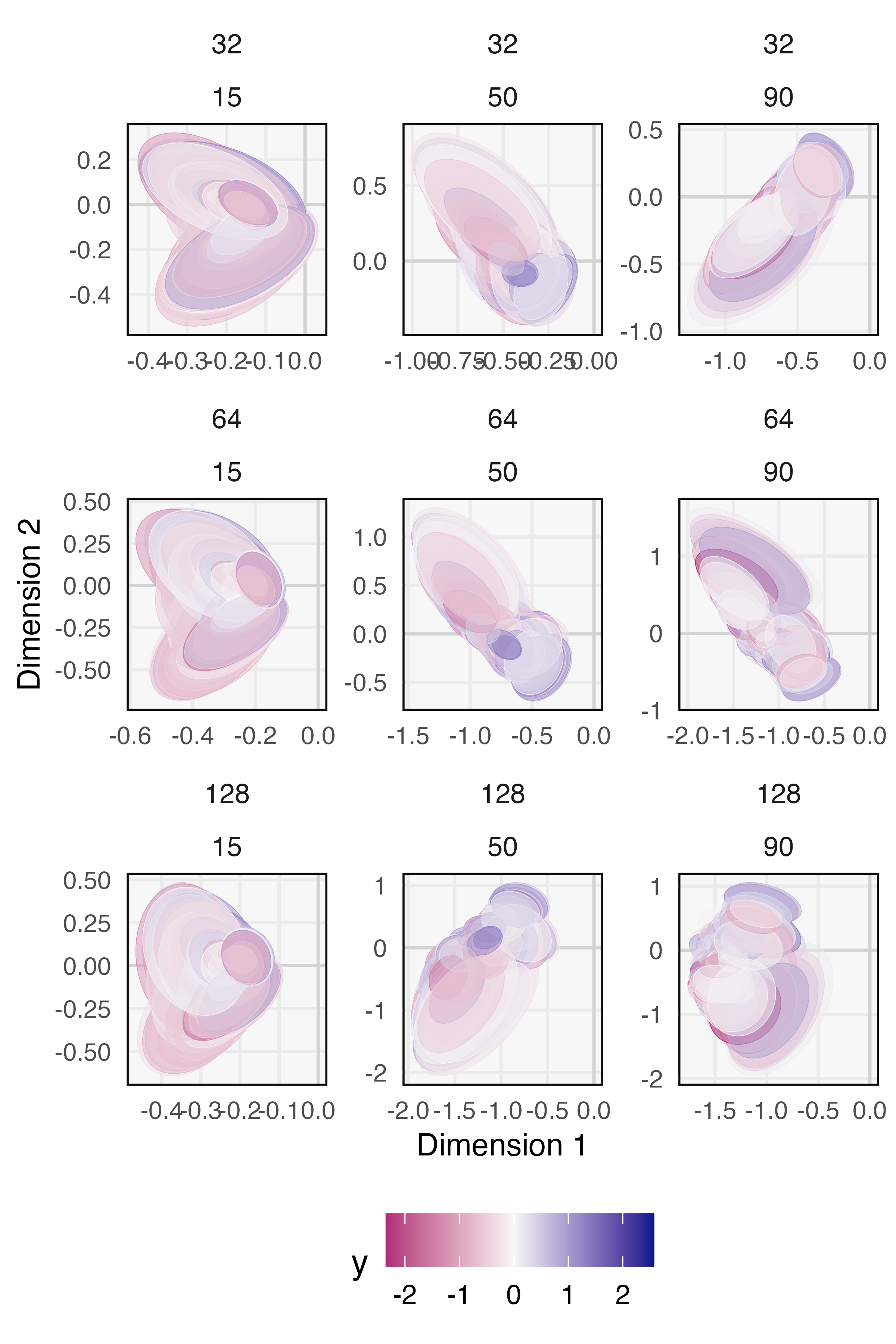}} 
    \subfloat{\includegraphics[width=0.5\textwidth]{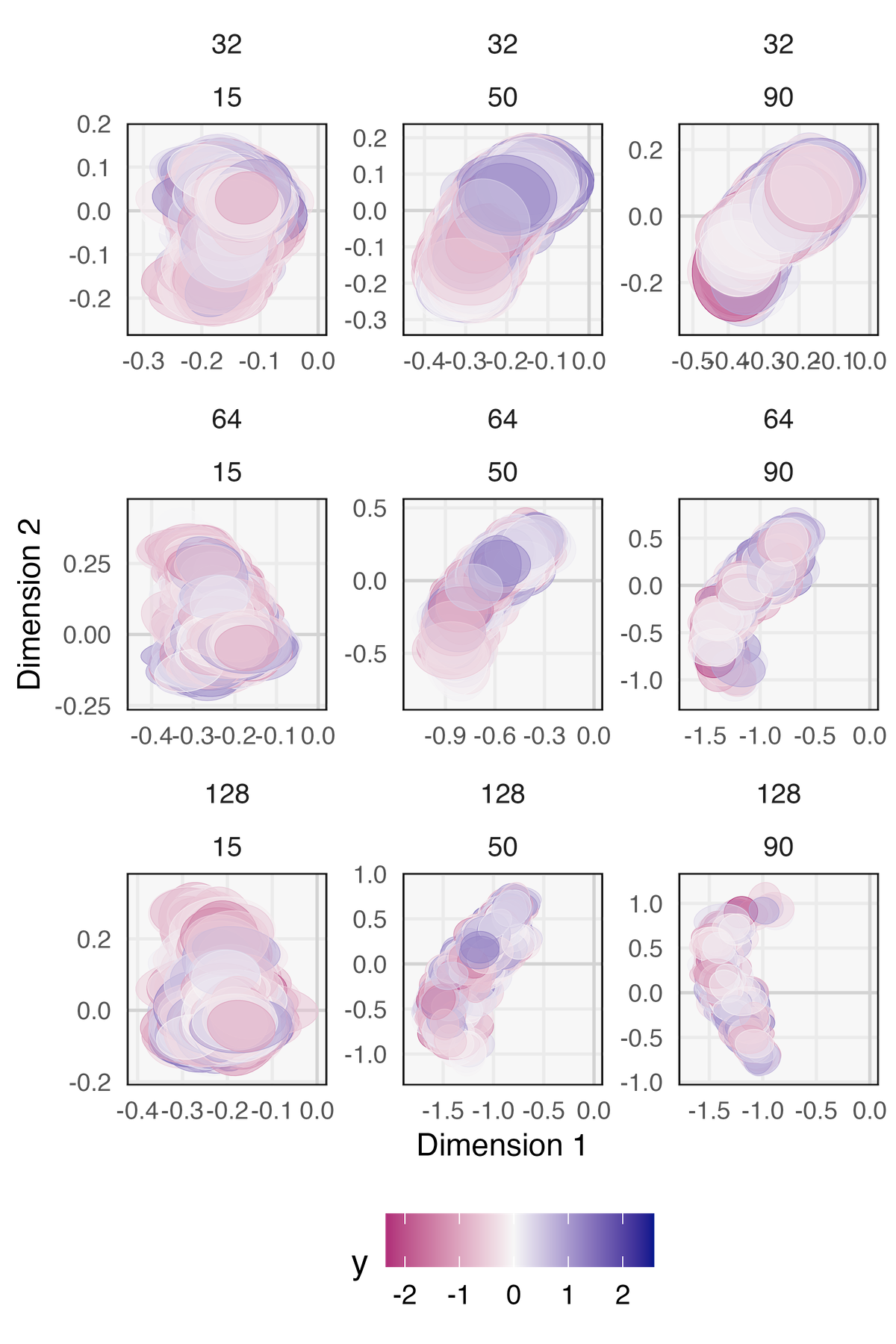}}
    \caption{The analog of Figure \ref{fig:simulation_projection_combined}, still shown on CNN projections, but for the nonparametric (left) and compromise (right) bootstrap approaches.}
    \label{fig:my_label}
\end{figure}

\begin{figure}
    \centering
    \subfloat{\includegraphics[width=0.33\textwidth]{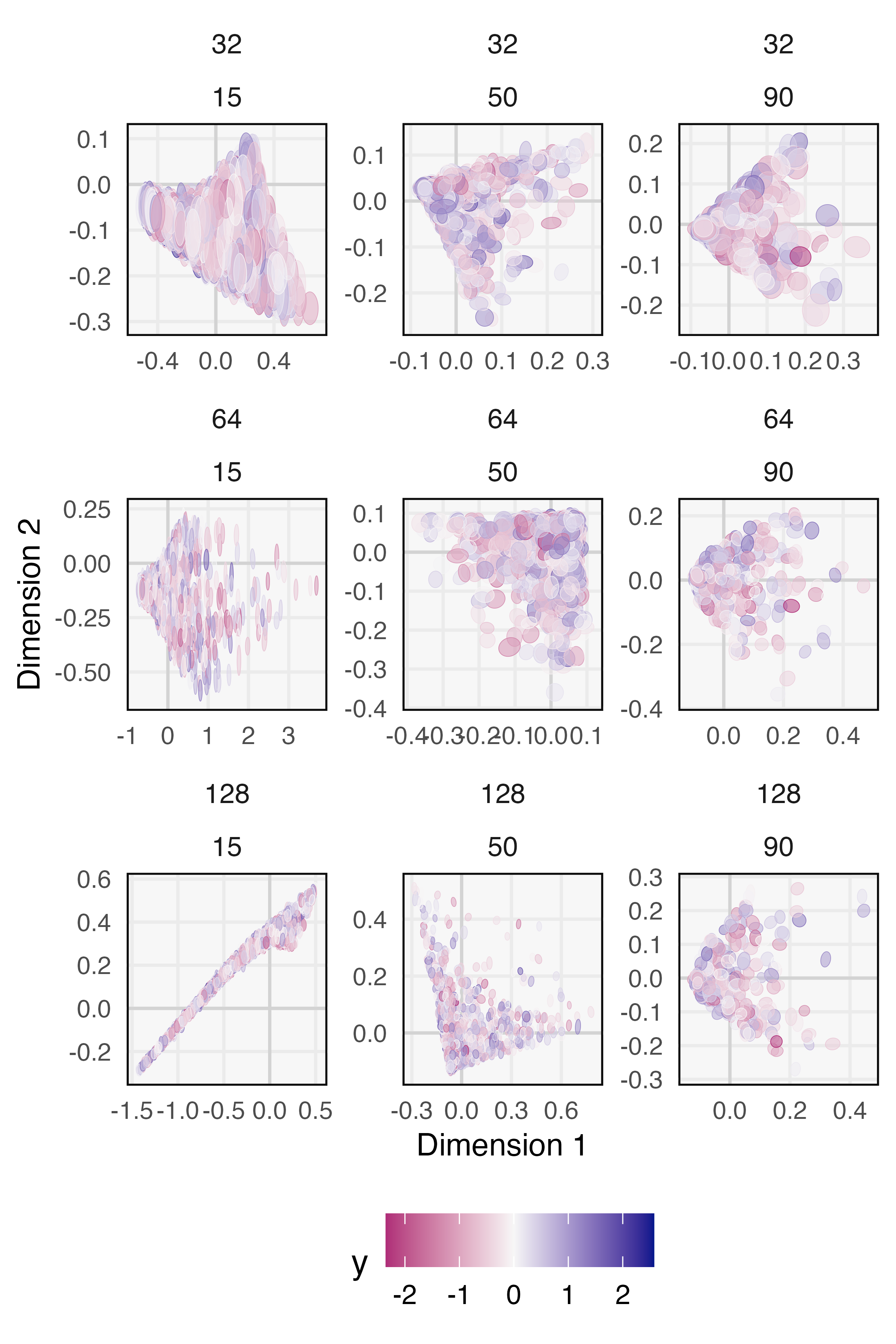}}
    \subfloat{\includegraphics[width=0.33\textwidth]{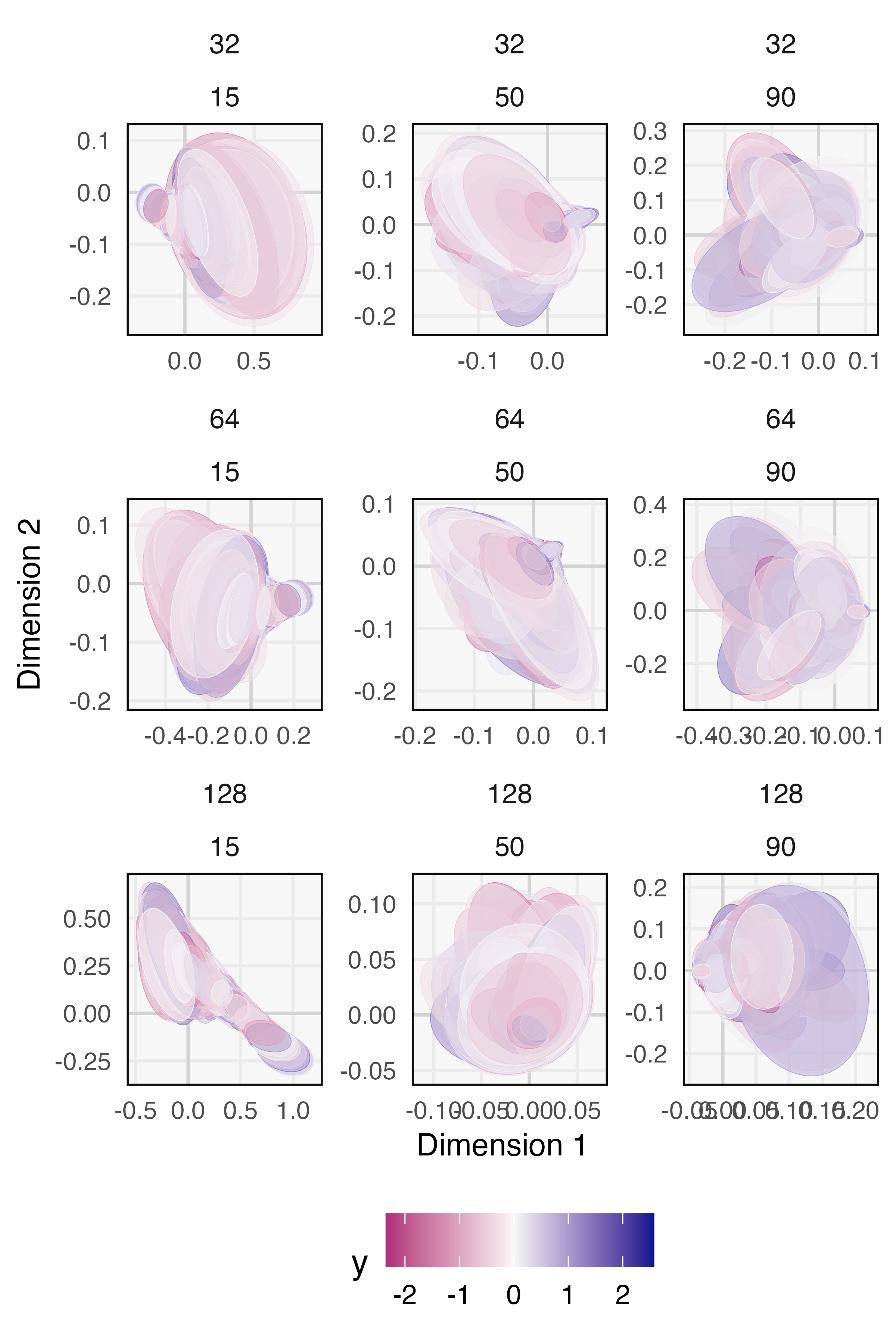}}
    \subfloat{\includegraphics[width=0.33\textwidth]{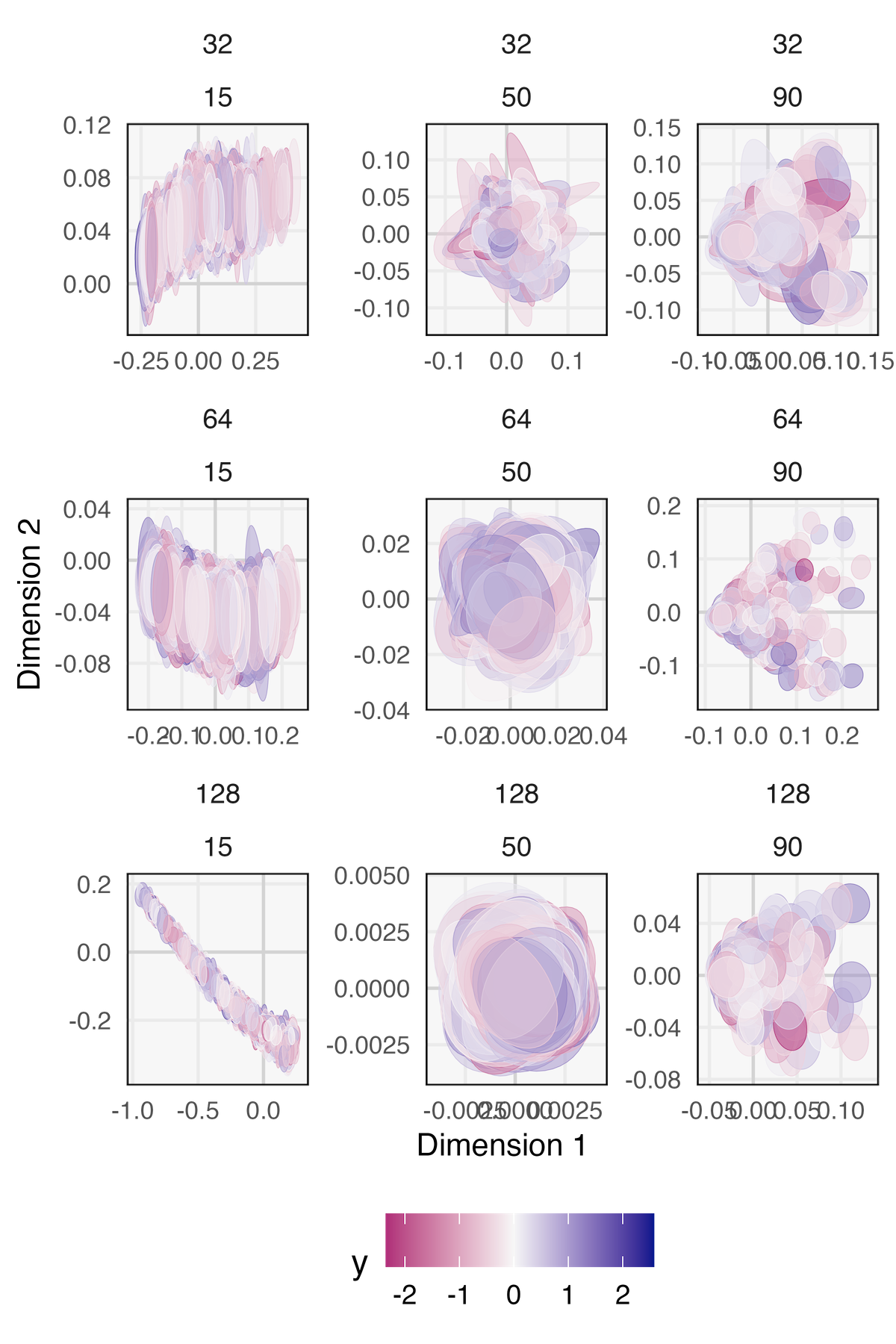}}
    \caption{The analog of Figure \ref{fig:simulation_projection_combined}, but for VAE models evaluated using the parametric (left), nonparametric (middle) and compromise (left) bootstrap approaches.}
    \label{fig:my_label}
\end{figure}

\begin{figure}
    \centering
    \subfloat{\includegraphics[width=0.5\textwidth]{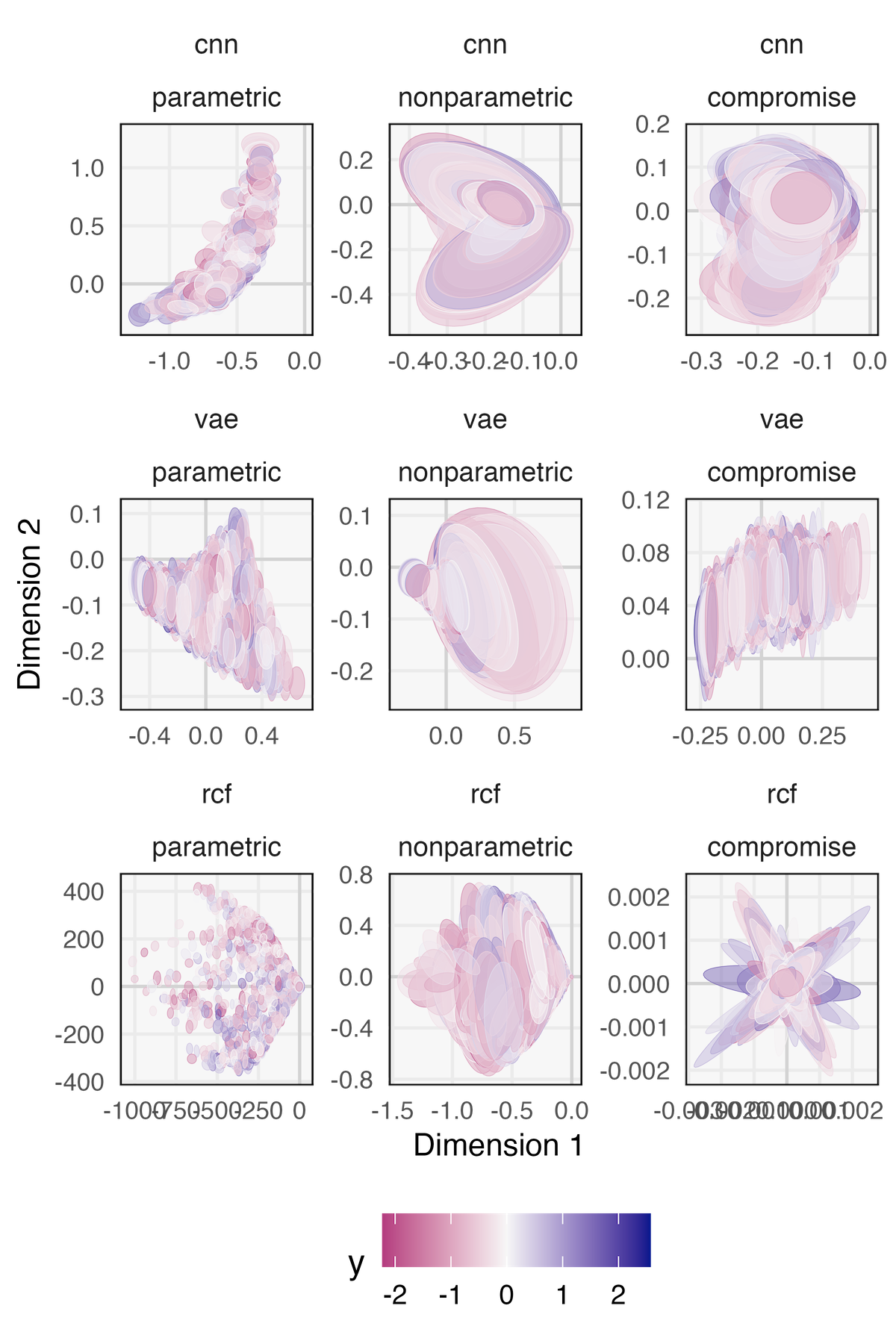}}
    \subfloat[]{\includegraphics[width=0.5\textwidth]{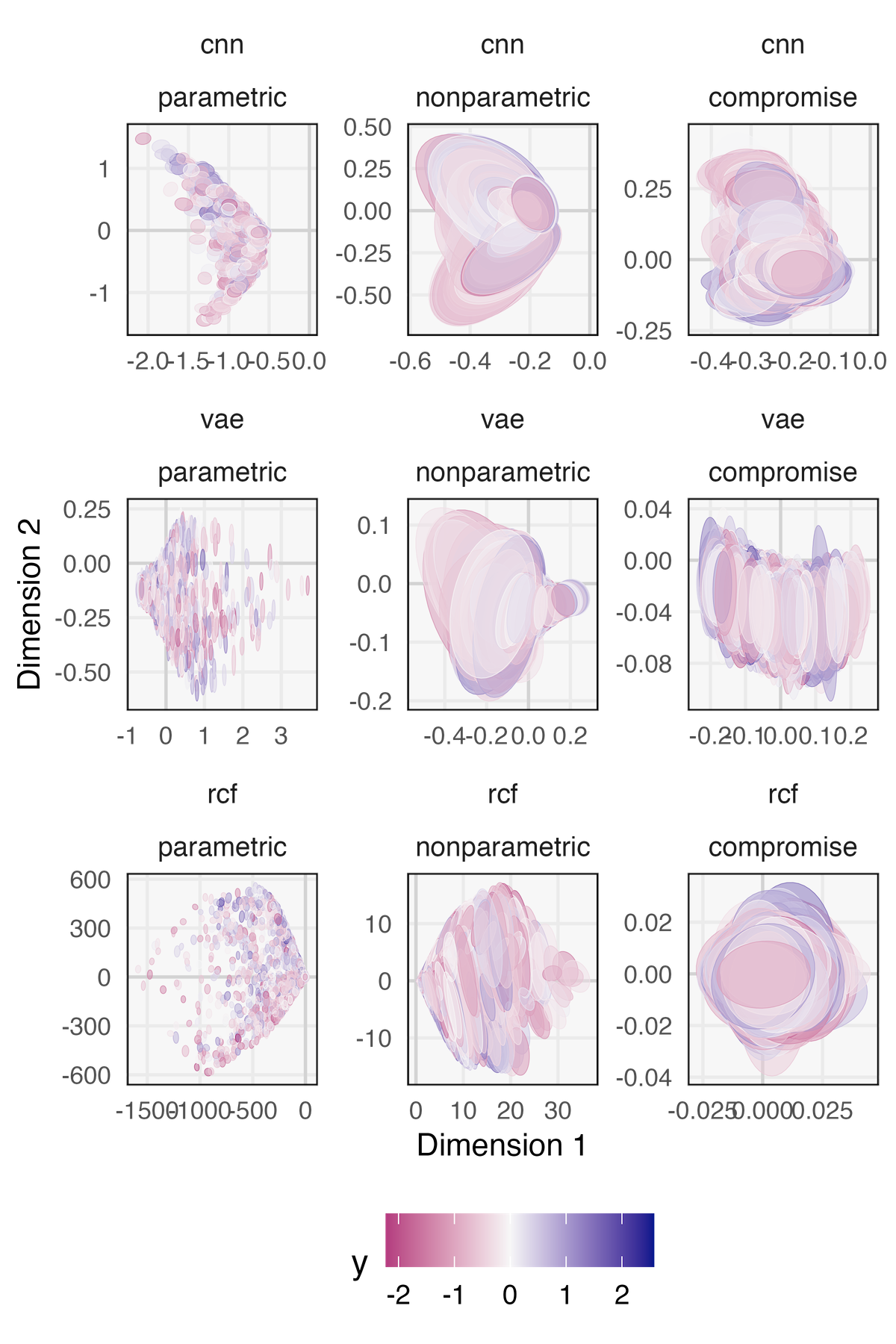}}
    \caption{Confidence areas corresponding to Figure \ref{fig:simulation_projection_combined}a, but for models with lower and average complexity, still using 15\% of data reserved for learning.}
    \label{fig:my_label}
\end{figure}

\begin{figure}
    \centering
    \subfloat{\includegraphics[width=0.33\textwidth]{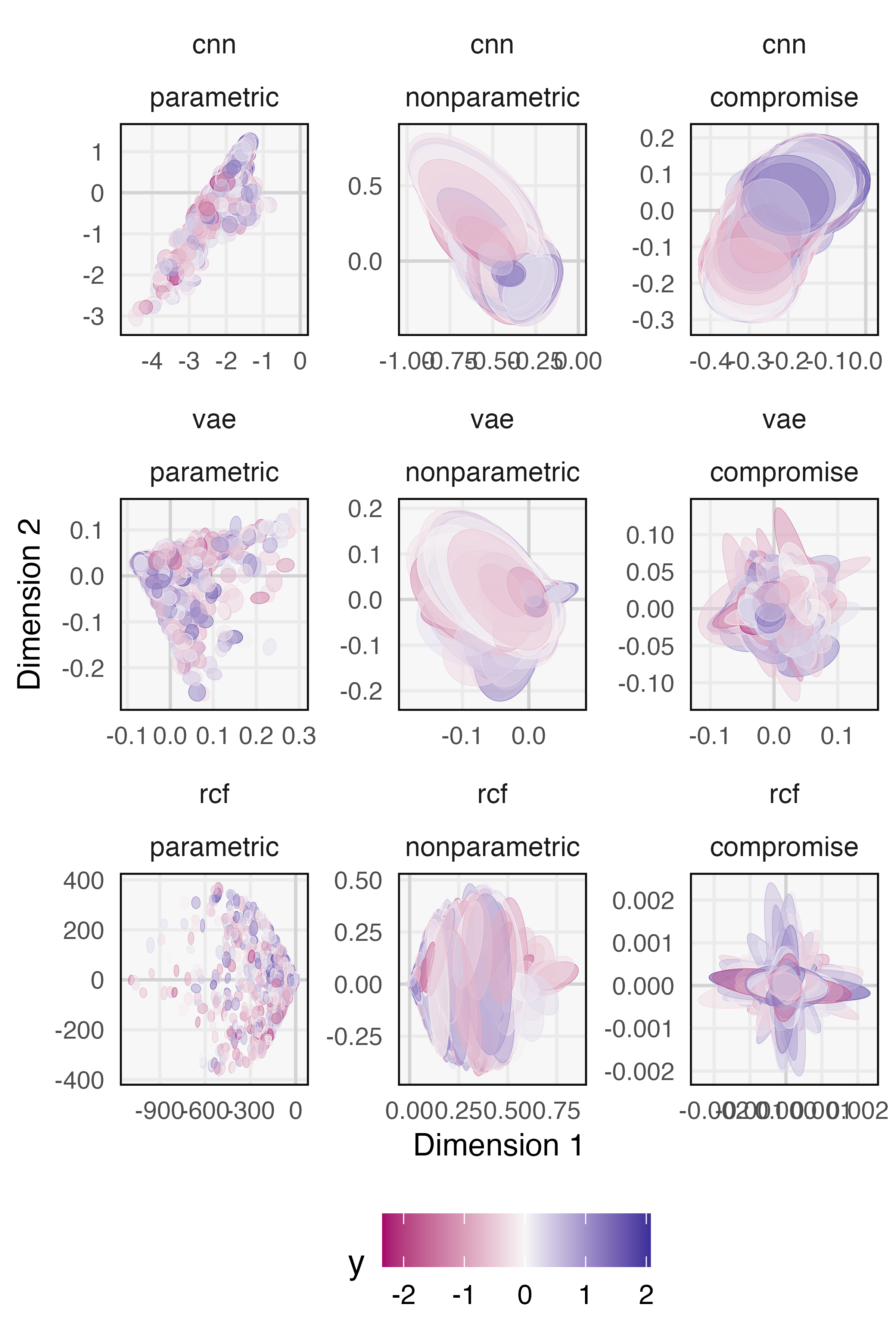}}
    \subfloat{\includegraphics[width=0.33\textwidth]{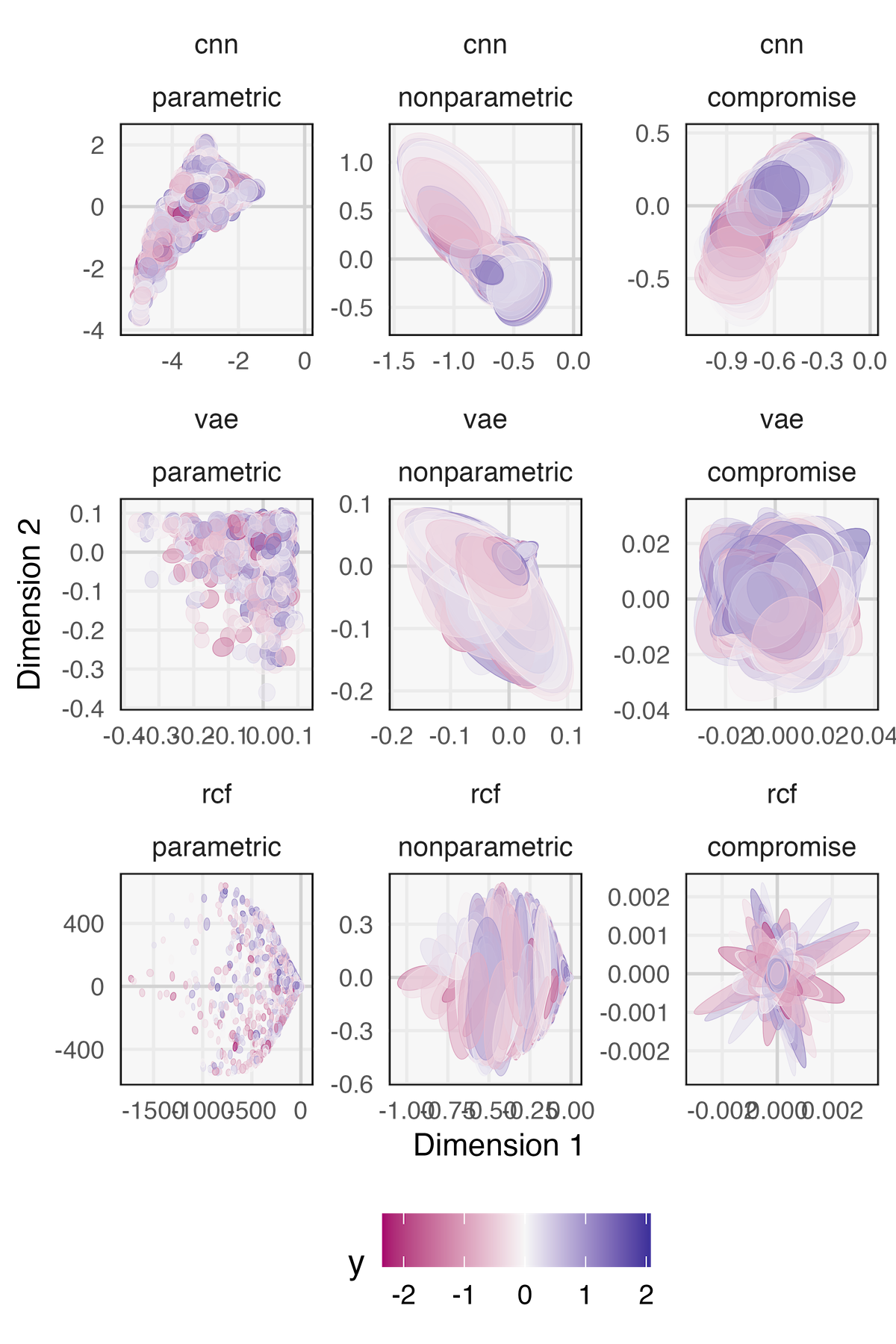}}
    \subfloat{\includegraphics[width=0.33\textwidth]{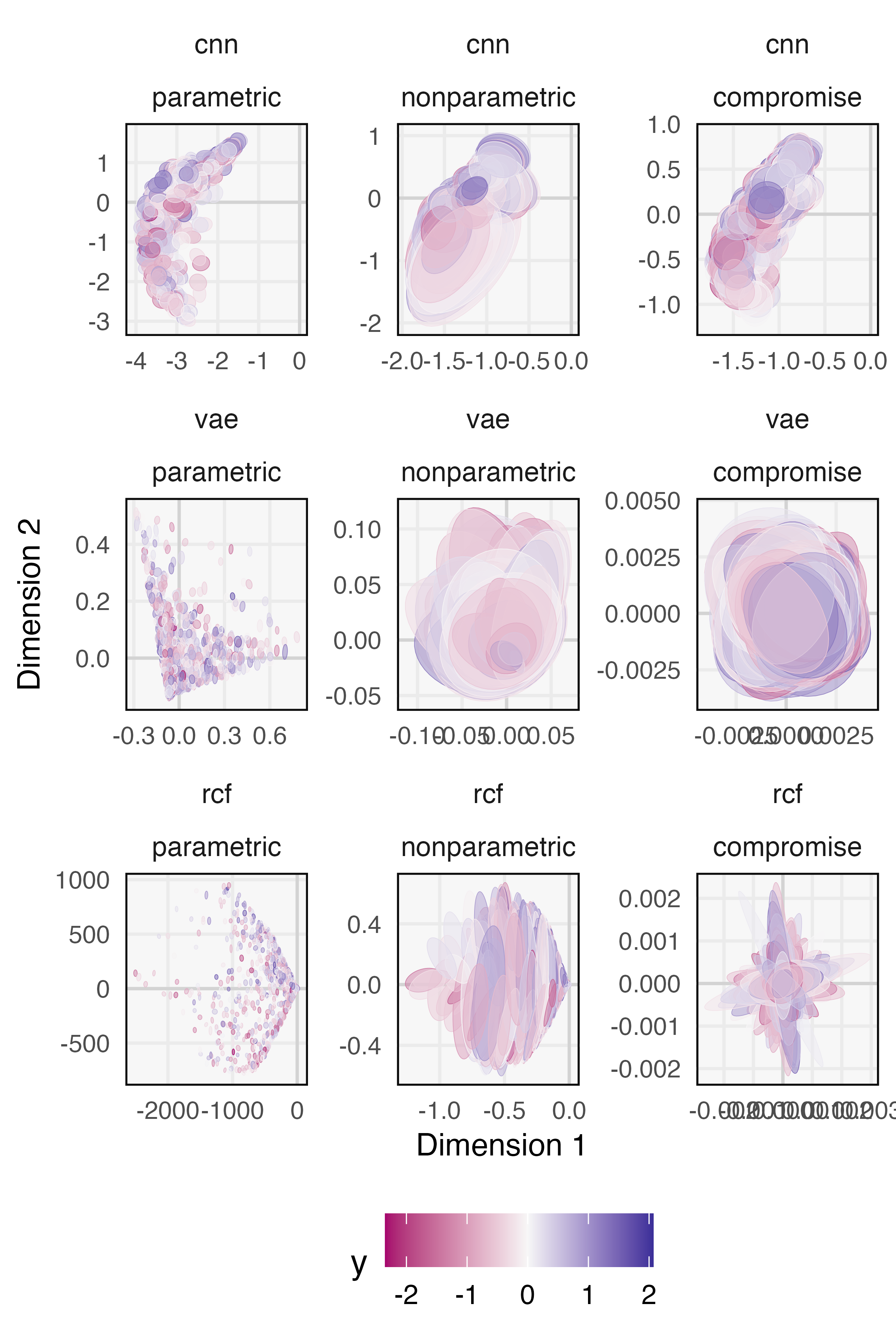}}
    \caption{Confidence areas analogous to Figure \ref{fig:simulation_projection_combined}a, but for models trained using 50\% of the data. The source images can be viewed in full resolution at the compendium repository.}
    \label{fig:my_label}
\end{figure}

\begin{figure}
    \centering
    \subfloat[]{\includegraphics[width=0.33\textwidth]{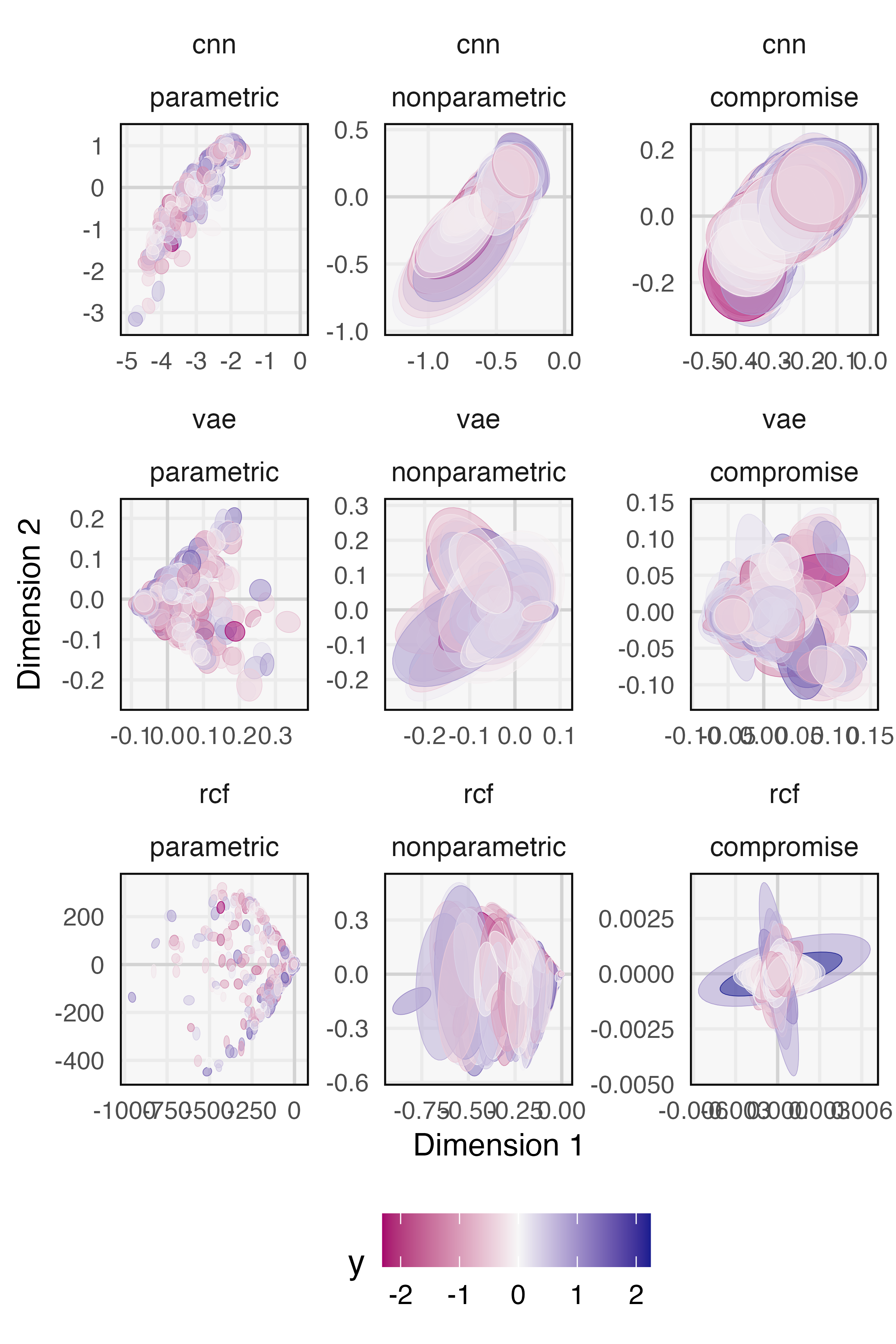}}
    \subfloat[]{\includegraphics[width=0.33\textwidth]{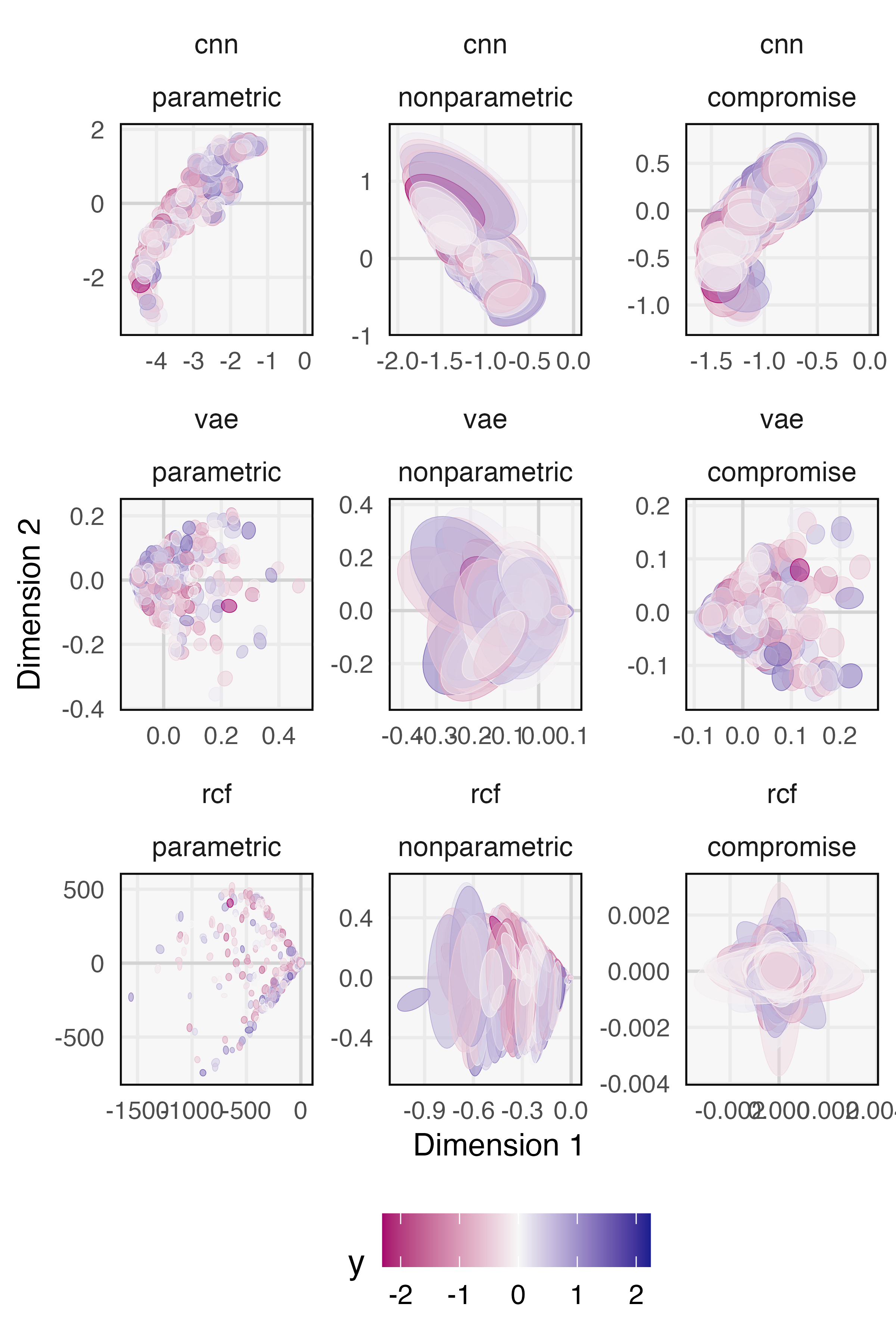}}
    \subfloat[]{\includegraphics[width=0.33\textwidth]{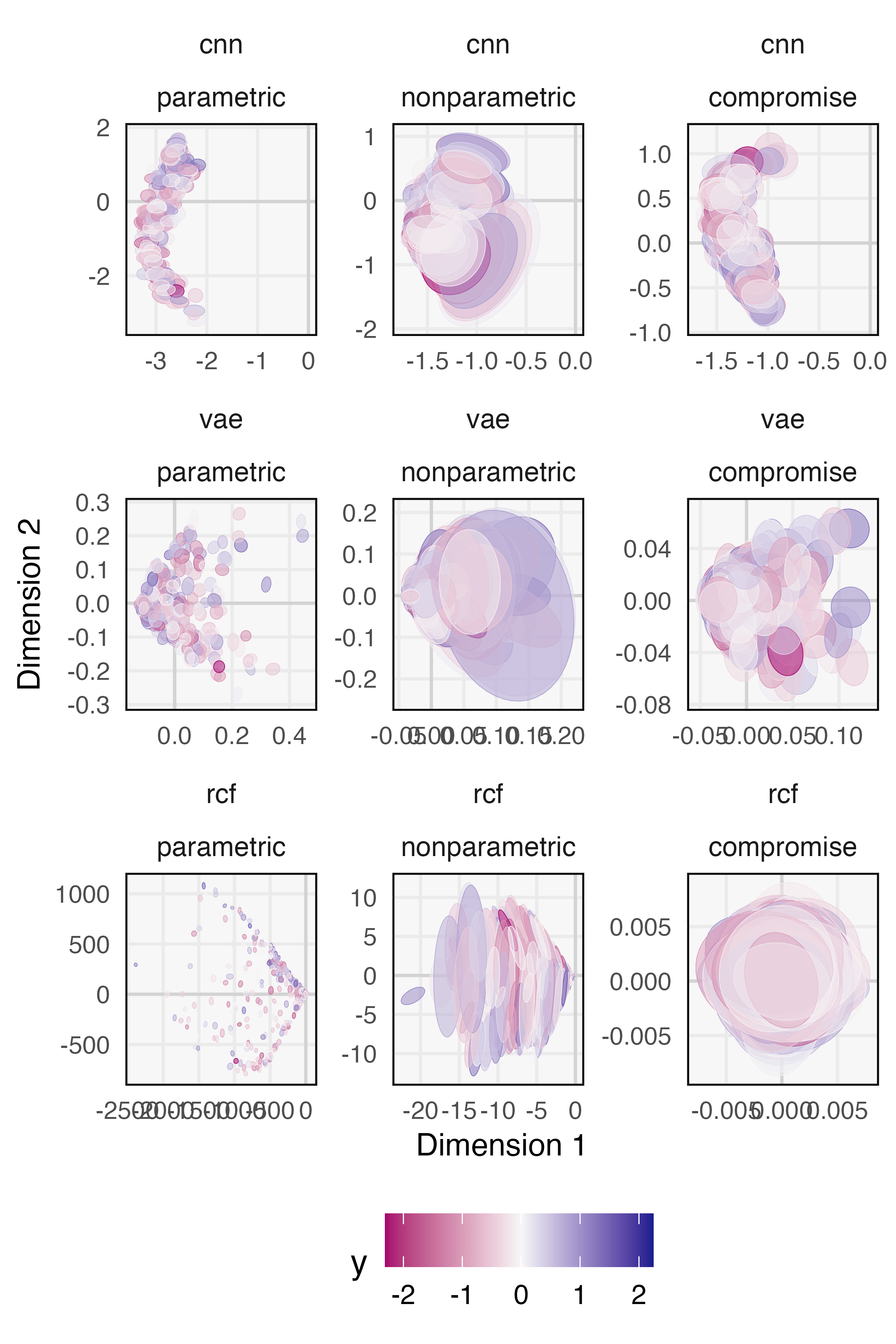}}
    \caption{Confidence areas analogous to Figure \ref{fig:simulation_projection_combined}a in the spatial point process simulation, but for models trained using 90\% of the data. The source images can be viewed in full resolution at the compendium repository.}
    \label{fig:my_label}
\end{figure}

\begin{figure}
    \centering
    \subfloat[]{\includegraphics[width=0.5\textwidth]{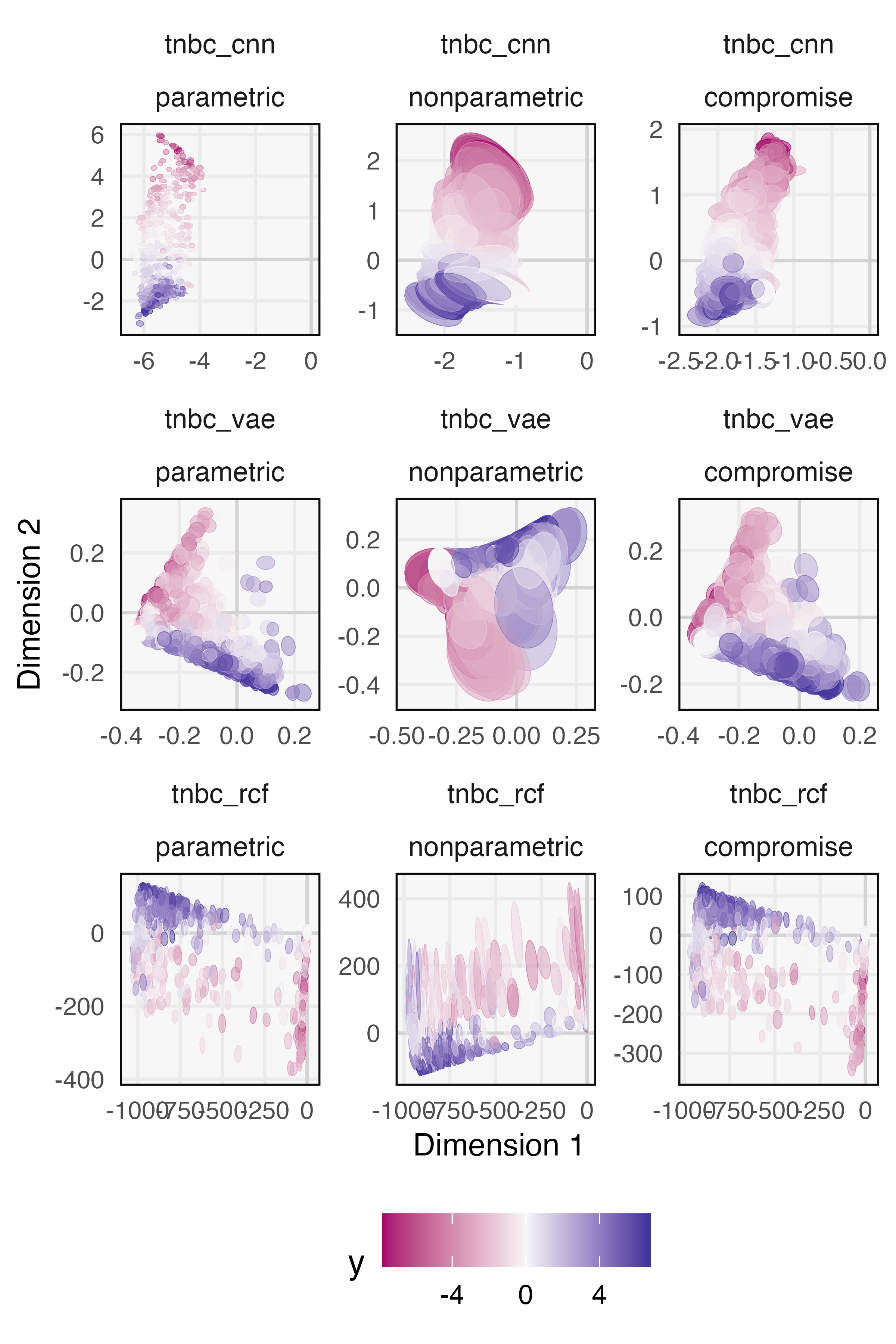}}
    \subfloat[]{\includegraphics[width=0.5\textwidth]{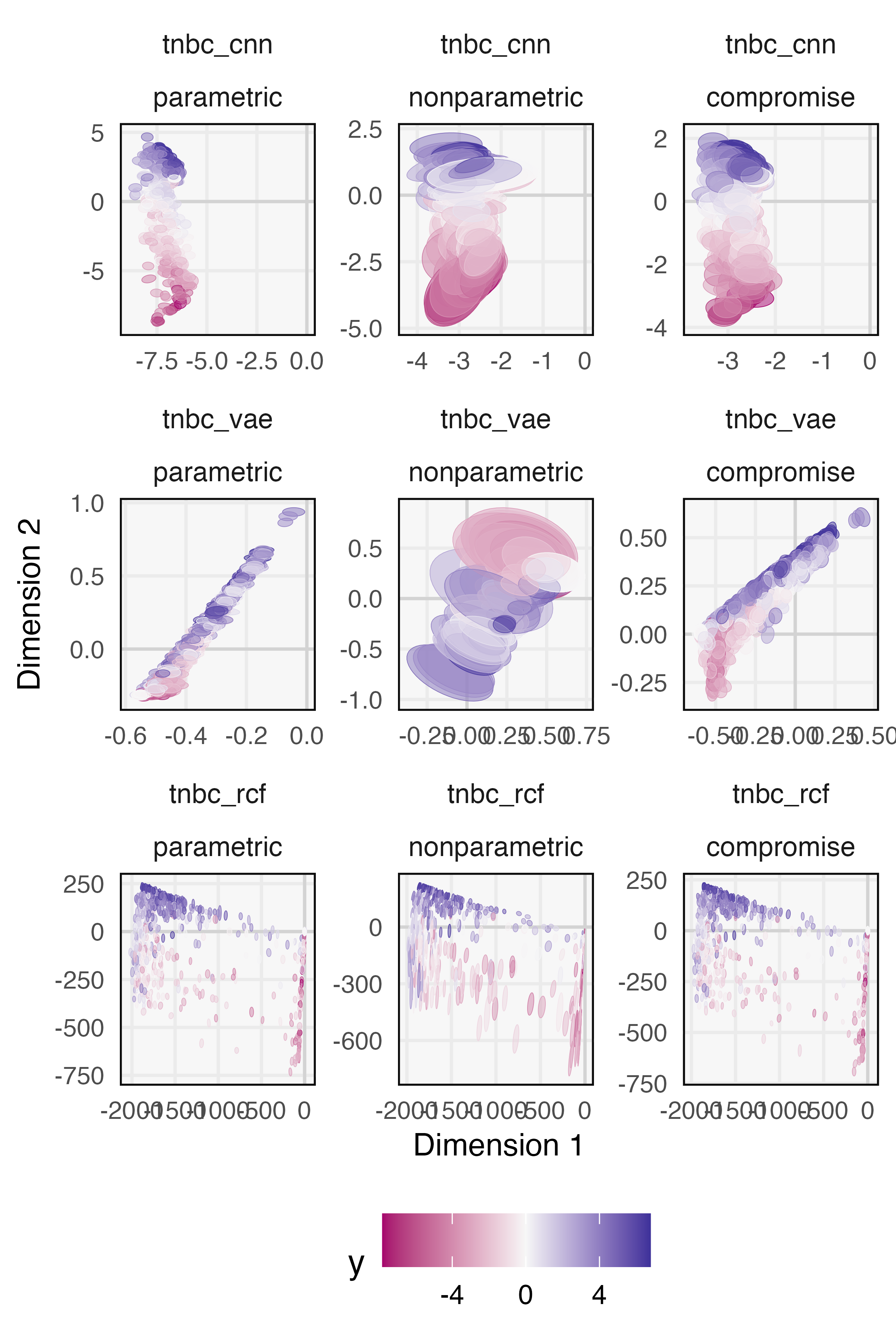}}
    \caption{Confidence areas analogous to Figure \ref{fig:64_coordinates} in the data analysis application, but for models with lower (left) and higher (right) complexity. The source images can be viewed in full resolution in the compendium repository.}
    \label{fig:my_label}
\end{figure}

\end{document}